\documentclass[11pt]{article}
\usepackage[dvipdfm]{graphicx}
\usepackage{amsmath,amsthm,amssymb}

\begin{document}
\title{Entropy production in 2D $\lambda \phi^4$ theory
\\
in the Kadanoff-Baym approach
}

\author{Akihiro Nishiyama\\
{\it Institute of Physics, University of Tokyo, 
Komaba, Tokyo 153-8902, Japan} }
\date{\today}
\maketitle

\abstract{%
We study non-equilibrium quantum dynamics of the single-component
scalar field theory in 1+1 space-time dimensions 
on the basis of the Kadanoff-Baym equation including the 
next-to-leading-order (NLO) skeleton diagrams.
As an extension of the non-relativistic case,
we derive  relativistic kinetic entropy
at the first order in the gradient expansion of the
Kadanoff-Baym equations.
The derived entropy satisfies the H theorem. 
Next we perform numerical simulations in spatially homogeneous
configurations to investigate thermalization properties of the system
by evaluating the system entropy.
We find that at later times the kinetic entropy increases approaching
the equilibrium value, although the limited time interval in the early
stage invalidates the use of it.
}

\section{Introduction}

Non-equilibrium quantum field theories provide a suitable framework
to investigate a large variety of topical problems 
in high energy particle physics,
astrophysics, cosmology, as well as condensed matter
physics\cite{BergesReview,Noneq}. 
In the context of heavy ion collision physics, the early time evolution
of the colliding system toward the quark-gluon plasma (QGP) state
has attracted a lot of theoretical interests for recent years.
Success of ideal hydrodynamic models for describing bulk properties
of the matter created at Brookhaven's Relativistic Heavy Ion Collider
(RHIC) seems suggesting that the 
produced system is strongly interacting and nearly thermalized
within a short time\cite{HeinzK} compared with perturbative analysis\cite{BMSS}.
There are various theoretical studies on the possibility for
this short time thermalization, some of which rely
on the instabilities in the plasma\cite{Weibel,Mrow1993,Arnold2003,Rebhan2005,Dumitru2007,Romat2006,NielsenO}, 
and some others include the 2-to-3 processes in
parton cascade simulations\cite{XuG2007}.

In the earliest stage of the high-energy nuclear collisions,
the system will be so dense that it would be more suitable to describe
the system in terms of the quantum field degrees of freedom
than in the particle basis.
As a first step of this approach toward the early time dynamics 
of the nuclear collisions,
we study here the non-equilibrium $\lambda \phi^4$
scalar field theory in 1+1 dimensions
on the basis of the Kadanoff-Baym (KB) equations.

As early as 1960s, based on a functional formulation of Luttinger and
Ward \cite{LW}, Baym and Kadanoff studied the Dyson-Schwinger equation
for the two-point function $G(x,y)$
\cite{BK}. 
Then Baym reformulated it in terms of variational principle,
introducing the so-called $\Phi$-derivable approximation
\cite{Baym,KB62}.
The functional $\Phi[G]$ is given by a truncated set of closed
two-particle irreducible (2PI) diagrams, and 
generates the driving terms of the equations of motion. 
The main virtue of this approximation is that 
the resulting KB equations conserve the energy and momentum of the
system.
This approach was extended to relativistic systems and
formulated using the path integral by Cornwall, Jackiw and Tomboulis in
\cite{CJT}. It  can be extended further to more general non-equilibrium
many-body systems based on the Schwinger-Keldysh real-time path integral
method\cite{Schwinger,Keldysh}.

In the last several years,
the real-time field dynamics has been newly investigated
by several authors. 
A seminal work was carried out by Danielewicz\cite{Danielewicz},
who for the first time studied the full KB equations
in the context of the heavy ion collisions
at non-relativistic energies.
He used a spatially homogeneous initial condition with
the non-spherical Fermi distribution for the nucleon momentum.
Thermalization problem 
in the relativistic $\lambda \phi^4$
scalar field theory in $1+1$ \cite{{AB},{BC}},
$2+1$ \cite{JCG} and $3+1$ \cite{{AST},{LM}} dimensions,
has been investigated with keeping the NLO skeleton diagrams in $\Phi$. 
Extension to the $O(N)$ theory 
at the next-to-leading order in $1/N$ expansion 
can be found in Ref.~\cite{Berges}. 
Importantly, all these analyses indicate that 
thermalization is achieved in course of the time evolution
of the system independently of the initial conditions.
The number distribution functions of the quasi-particles
are found to approach the Bose distribution.

The approach to the equilibrium state will be quantitatively
characterized if system entropy can be introduced properly.
In fact it is an open question how to choose the gross variables
and define the corresponding entropy of the
system in general non-equilibrium situations.
There is no entropy production in fully microscopic calculations.
We use the variable $G(x,y)$ in the KB approach.
In the non-relativistic case, the kinetic entropy is introduced 
at the first order of the gradient expansion
in Refs.~\cite{IKV4}
and \cite{Kita}
\footnote{However, their expressions are different from each other 
in the higher order terms of the skeleton expansion.}.
To our knowledge, the  entropy production has not ever been
estimated in the relativistic KB dynamics. 
Here we shall extend the entropy to the relativistic case in the first
order of the gradient expansion.
This will provide us, for example, of a criterion how much each
microscopic process contributes to thermalization of the system.

The KB equations deal with 
the evolution of the field $\langle \phi \rangle$ 
and the two-point function $G(x,y)$,
and effectively contain particle number
changing processes such as $1\leftrightarrow 3$
even in the NLO,
if interpreted in the particle basis.
In contrast, the Boltzmann equation includes 
only the $2\leftrightarrow 2$ scattering processes 
to this order, which preserves the total particle number.
This difference should be reflected in the behavior of the system 
evolution, especially in the entropy production.
We expect that this aspect of the KB equations
is important to understand the possibility of 
the rapid thermalization.

For demonstration, 
we shall numerically solve the non-equilibrium dynamics of 
$\lambda \phi^4$ theory on the basis of the KB equations.
In order to reduce the numerical cost,  
we restrict our simulations
to the spatially uniform case without the mean field
$\langle \phi \rangle =0$ in $1+1$ dimensions.
We start the simulations with the non-thermal initial conditions,
and show the time evolution of the system through
the particle number distribution functions, the energy content, the
entropy production 
and so on.

This paper is organized as follows.
In Sec. 2 we briefly review the formulation of the KB
equation for the relativistic scalar field theory,
using the Schwinger-Keldysh formalism and 2PI effective action
in the NLO.
Next, we present the derivation of the entropy for the relativistic KB
equations in the first order in the gradient expansion in Sec. 3, 
which is the main theoretical part of this work. 
The expression for the entropy is found to be a natural extension
of the non-relativistic one given in \cite{{IKV4},{Kita},{IKV},{IKV2}} 
in the local approximation. This entropy satisfies the H-theorem.
In Sec. 4 we show the numerical simulations of the KB equations.
The particle number distribution and the entropy of the system are 
calculated in terms of the numerical solutions of the KB equations.
Finally Secs. 5 and 6 are devoted to discussions and summary of this study.

\section{Kadanoff Baym equation}

We briefly review the derivation of Kadanoff-Baym equation 
and fix our notations\cite{BergesReview}.
For the scalar field theory 
$\mathcal{L}=\frac{1}{2}\partial_\mu \phi
\partial^\mu \phi -\frac{1}{2}m^2\phi^2 -\frac{1}{4!}\lambda \phi^4$,
the 2PI effective action with vanishing mean field $\langle \phi \rangle=0$ (unbroken phase) 
is written as  
\begin{eqnarray}
\Gamma[G] &=&
\frac{i}{2} {\rm Tr} {\rm ln} \left( G \right) ^{-1} 
+ \frac{i}{2} G_0^{-1} G +\frac{1}{2} \Phi[ G]\, .
\label{Action}
\end{eqnarray}
Here $iG_0^{-1}(x,y)=-(\partial^2_x+m^2)\delta_{\cal C}(x-y)$
is the free Green's function 
and $G$ is the full Green's function, both of which are defined on the
closed time path $\mathcal{C}$.
The functional $\Phi[G]$ in (\ref{Action}) is generally a sum of all
possible 2PI graphs written in terms of $G$.
A graph is called 2PI when
it remains connected upon cutting two Green's function lines.

The stationary condition for the effective action (\ref{Action})
\begin{eqnarray}
\frac{\delta \Gamma}{\delta G}=0 
\label{sc}
\end{eqnarray}
gives rise to the Schwinger-Dyson equation for
the Green's function $G(x,y)$
\begin{eqnarray}
G^{-1}(x,y) = 
G^{-1}_{0}(x,y) - \Sigma(x,y)
\label{S-D}
\end{eqnarray}
with the proper self-energy defined as
$\Sigma= i{\delta \Phi[G] } /{\delta G}$.
The self-energy is divided into the local and the non-local part
$\Sigma= \Sigma_{\mathrm{loc}}+\Sigma_{\mathrm{nonl}}$.
The $\Sigma_{\mathrm{loc}}$ contributes to the effective mass
while the $\Sigma_{\mathrm{nonl}}$
induces the mode-coupling between the different wavenumbers.
The 2PI effective action should be invariant under the symmetry
transformations of the system.
Although we need to approximate the functional $\Phi[G]$ in practical
applications, any truncation of $\Phi[G]$ which preserves the symmetry property
gives the equations of motion consistent with the corresponding
conservation laws\cite{{BK},{Baym}}.

It is very useful to decompose the two-point function $G(x,y)$ into
two real functions, 
the statistical function $F(x,y)$ and the spectral function $\rho(x,y)$
defined, respectively, as
\begin{eqnarray}
F(x,y)= 
\frac{1}{2} \left \langle  \left \{ \phi(x), \phi(y) \right \}
\right \rangle =  
\frac{1}{2} \left[ G^{21}(x,y) + G^{12}(x,y )  \right ] 
\end{eqnarray}
and\begin{eqnarray}
\rho(x,y)= 
i \left \langle  \left[ \phi(x), \phi(y) \right] \right \rangle 
= 
i \left[ G^{21}(x,y) - G^{12} (x,y )  \right ]
\; ,
\label{eq:rho}
\end{eqnarray}
where $\langle \cdots \rangle $ represents the expectation value taken 
over a certain initial density matrix.
The indices 1 and 2 specify the branch of the contour $\mathcal{C}$ in 
the Schwinger-Keldysh formalism.
The function $F$ is called the statistical function because 
it turns out to be the Bose distribution function
in the equilibrium state.
The Schwinger-Dyson equation (\ref{S-D}) can be equivalently
rewritten in terms of $F(x,y)$ and $\rho(x,y)$ as
coupled integro-differential equations
\begin{eqnarray}
(\partial^2+m^2+\Sigma_{\rm loc}(x))F(x,y)
&=&
 \int _{t_0}^{y^0} dz\Sigma_F (x,z) \rho (z,y)
-\int _{t_0}^{x^0} dz\Sigma_\rho (x,z) F(z,y) \; ,
\label{kbf}
\\
(\partial^2+m^2+\Sigma_{\rm loc}(x))\rho(x,y)
&=&
-\int ^{x^0}_{y^0} dz\Sigma_\rho (x,z) \rho (z,y)\;,
\label{kbr}
\end{eqnarray}
where $t_0$ is the initial time.
Note that the non-local self-energy has been re-expressed similarly
as
\begin{eqnarray}
\Sigma_{F}(x,y)
&=&
\frac{1}{2}
\left[
\Sigma^{21}_{\rm nonl}(x,y)+ \Sigma^{12}_{\rm nonl}(x,y)
\right ]
\; ,
\\
\Sigma_{\rho}(x,y)
& =&
i\left[\Sigma^{21}_{\rm nonl}(x,y)- \Sigma^{12}_{\rm
  nonl}(x,y)\right ]
\, .
\end{eqnarray}
The set of equations (\ref{kbf}) and (\ref{kbr}) is 
called the Kadanoff-Baym equation, which is the 
{\it two-time} formalism and
describes the time evolution of the system from 
a certain initial configuration for $F$ and $\rho$.
Note that at each time step
the spectral function $\rho$ must satisfy the 
conditions following from the commutation relations:
\begin{eqnarray}
\rho (x,y) |_{x^0 \rightarrow y^0}&=&0
\; ,
\nonumber \\
\partial_{x^0} \rho (x,y)| _{x^0 \rightarrow y^0}& =&
\delta^{d}({\bf x-y}) 
\; ,
\nonumber \\ 
 \partial_{x^0}\partial_{y^0} \rho(x,y)|_{x^0 \rightarrow y^0}  &=& 0
\; .
\label{eq:rhocon}
\end{eqnarray}

Importantly,
Eqs. (\ref{kbf}) and (\ref{kbr}) are non-local in time
due to the so-called memory integrals appearing on the RHS.
In other words,
the evolution is non-Markovian depending on the evolution history
in the past.
In many stable systems, however, 
the integrand of the memory integral dies
away exponentially and the macroscopic time scale is separated from
the microscopic one.

It is instructive to consider the case of a uniform equilibrium state
with a small value for the self-energy $\Sigma_\rho(p^0,p)$. 
Then we find that the spectral function $\rho$ turns out to be the
Breit-Wigner form
(See Sec.~\ref{sec:entropy}),
\begin{eqnarray}
\rho(p^0,p)
= \frac{-\Sigma_\rho}
 {\left[(p^0)^2-\Omega_{\bf p}^2\right]^2-\Sigma_\rho^2/4}
\rightarrow 
2\pi i \epsilon(p^0) \delta((p^0)^2-\Omega_{\bf p}^2)
\; ,
\label{eq:rfree}
\end{eqnarray}
where $\Omega_{\bf p}^2={\bf p}^2+m^2+{\rm Re}\Sigma_{R}$
is the single particle energy including the mean-field effect.
The arrow denotes the quasi-particle limit $\Sigma_\rho \to 0$.
In this limit the $\rho$ becomes a delta-function
and for thermal equilibrium the statistical function $F$ reduces to
the Bose distribution 
\begin{eqnarray}
F(p^0,{ p})=
2\pi \delta((p^0)^2- \Omega_{\bf p}^2) 
\left(\frac{1}{2}+\frac{1}{e^{\beta p^0}-1}\right)
\; .
\label{eq:ffree}
\end{eqnarray}

\begin{figure}
\begin{center}
\includegraphics[width=160pt,height=65pt]{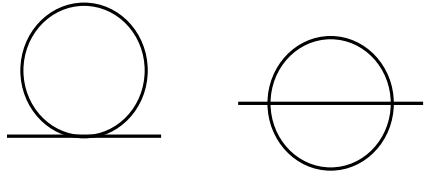}
\end{center}
\caption{Tadpole and sunset diagrams. }
\label{fig:tadsun}
\end{figure}

In this paper we
restrict ourselves to the spatially homogeneous situation.
From the translational invariance, the statistical function
$F(x,y)=F(x^0,y^0,{\bf x-y})$ and 
the spectral function $\rho(x,y)=\rho(x^0,y^0,{\bf x-y})$
can be Fourier transformed to $F(x^0,y^0; {\bf p})$ and 
$\rho(x^0,y^0; {\bf p})$. 
Then KB equations are simplified in the momentum space as
\begin{subequations}
\begin{align}
(\partial_0^2+{\bf p}^2+m^2+\Sigma_{\rm tad}(x^0))F(x^0,y^0;{\bf p})
=&
 \int _{t_0}^{y^0} dz^0\Sigma_F (x^0,z^0;{\bf p})  \rho
(z^0,y^0;{\bf p}) \nonumber \\
&   -\int _{t_0}^{x^0} dz^0\Sigma_\rho(x^0,z^0;{\bf p}) F(z^0,y^0;{\bf p})
\; ,
 \\
(\partial_0^2+{\bf p}^2+m^2+\Sigma_{\rm tad}(x^0))\rho(x^0,y^0,{\bf p}) 
=&
 -\int ^{x^0}_{y^0} dz^0\Sigma_\rho (x^0,z^0;{\bf p}) 
 \rho(z^0,y^0;{\bf p})
\; .
\end{align}
\label{K-B}
\end{subequations}
Regarding the functional $\Phi[G]$, we approximate it 
with the skeleton diagrams obtained at the next-leading order in $\lambda$.
The self-energy $\Sigma$ then becomes the sum of
the local tadpole diagram and the nonlocal sunset diagram
(Fig.~\ref{fig:tadsun}):
\begin{eqnarray}
&&
\Sigma_{\rm loc}(x)=\Sigma_{\rm tad}(x) =\frac{\lambda}{2} F(x,x) ,
\label{sigtad}
\\
&&
{\Sigma^{ab}_{\rm sun}}  (x,y)= -\frac{\lambda^2}{6} {G^{ab} (x,y) }^3
\; 
\label{signonl}
\end{eqnarray}
where indices $a,b$ denote the branch 1 and 2 of Schwinger-Keldysh contour ${\cal C}$. 
Furthermore the nonlocal part is divided into $\Sigma_{F}$
and $\Sigma_{\rho}$
and written explicitly in terms of $F$ and $\rho$ as
\begin{eqnarray}
\Sigma_F(x^0,z^0;{\bf p})
&=&
-\frac{\lambda^2}{6} \int \!\! \frac{d^d k}{(2\pi)^d} 
\frac{d^d q}{(2\pi)^d}  F(x^0,z^0;{\bf p-k-q} )
\nonumber \\
&& 
\times \Bigg[ 
F(x^0,z^0;{\bf k})F(x^0,z^0;{\bf q})
-\frac{3}{4} \rho(x^0,z^0;{\bf k})\rho (x^0,z^0;{\bf q})
 \Bigg]
\; ,
\label{sigf}
\\
\Sigma_\rho(x^0,z^0;{\bf p})
&=&
 -\frac{\lambda^2}{2} \int \!\! \frac{d^d k}{(2\pi)^d} \frac{d^d q}{(2\pi)^d} 
 \rho(x^0,z^0;{\bf p-k-q} ) 
\nonumber \\
&&
\times \Bigg[ F(x^0,z^0;{\bf k})F(x^0,z^0;{\bf q})
-\frac{1}{12} \rho(x^0,z^0;{\bf k})\rho (x^0,z^0;{\bf q}) \Bigg]
\; .
\label{sigr}
\end{eqnarray}
We solve these KB equations (\ref{K-B}) with the self-energy functions
(\ref{sigtad}), (\ref{sigf}) and (\ref{sigr}) numerically 
in Sec.~\ref{sec:numerics}.

We need to specify the initial condition for $\rho$ and $F$ 
at $x^0=y^0=t_0$ in order to solve this evolution equations.
For the spectral function $\rho$ it is fixed by the commutation relation
as given in Eqs.~(\ref{eq:rhocon}).
For the statistical function $F$, we choose to set the initial conditions of
the following functional form
\begin{eqnarray}
 \ F(x^0,y^0;{\bf p})\big |_{x^0=y^0=t_0}
&=&
\frac{1}{\omega({\bf p})}\left(n_ {\bf p}+\frac{1}{2} \right) 
\; , 
 \\
\partial_{x^0} F(x^0,y^0;{\bf p}) \big |_{x^0=y^0=t_0}
&=&
0 
\; , 
\\
\partial_{x^0}\partial_{y^0} F(x^0,y^0;{\bf p})\big |_{x^0=y^0=t_0}
&=&
{\omega({\bf p})}\left(n_{\bf p}+\frac{1}{2} \right)     
\; , 
\end{eqnarray}
where $\omega({\bf p})^2={\bf p}^2+m^2 $
and $n_{\bf p}$ is a function we can freely specify.
This form is assumed in analogy with the equilibrium solution 
in the quasi-particle limit.

At later times in course of the evolution, we 
{\it define} the particle number distribution $n_{\bf p} (X^0)$
and the frequency $\tilde \omega_{\bf p}(X^0)$
in terms of $F(x^0,y^0;{\bf p})$ 
\cite{{AB},{JCG},{AST},{LM},{BergesReview}}
\begin{eqnarray}
&&
n_{\bf p} (X^0) +\frac{1}{2} =
\left [ 
\partial_{x^0}\partial_{y^0} 
 F(x^0,y^0;{\bf p} ) \Big |_{x^0=y^0=X^0} F(X^0,X^0;{\bf p} )
-\left(\partial_{x^0}F(x^0,y^0;{\bf p}) \Big |_{x^0=y^0=X^0}\right)^2
\right ]^{1/2}
\; ,
\label{eq:np}
\\
&&
\tilde \omega_{\bf p}(X^0)= 
\left [
\frac{\partial_{x^0}\partial_{y^0} F(x^0,y^0;{\bf p} ) \big |_{x^0=y^0=X^0}}
     { F(X^0,X^0;{\bf p} )}
\right ] ^{1/2}
\; .
\label{eq:op}
\end{eqnarray}
Strictly speaking,
these definitions (\ref{eq:np}) and (\ref{eq:op}) are
valid only when the quasi-particle picture works well.
Nevertheless, we expect that these quantities are good estimators to
characterize the behavior of the system evolution.  
The system is expected to have a quasi-particle spectrum 
for a sufficiently small coupling $\lambda$ 
as shown in $1+1$ \cite{AB,BC}, $2+1$ \cite{JCG} and
$3+1$\cite{{AST},{LM}} dimensions.

Before proceeding to the next section let us compare the KB equations
with the Boltzmann equation in 1+1 dimensions.
In a homogeneous system
the Boltzmann equation becomes
\begin{eqnarray}
\Omega_{\bf p} \frac{\partial}{\partial t} n_{\bf p}(t) 
&=&
\frac{\lambda^2 }{4} \int \!\! \frac{d^{d}p_1}{(2\pi)^{d} } 
\frac{d^{d}p_2}{(2\pi)^{d} } 
\frac{d^{d}p_3}{(2\pi)^{d} } 
\frac{1}{8\Omega_{\bf p_1}\Omega_{\bf p_2}\Omega_{\bf p_3}}   
\nonumber \\
&&\times  \left [(1+n_{\bf p_3})(1+n_{\bf p})n_{\bf p_1}n_{\bf p_2}-
n_{\bf p_3}n_{\bf p}(1+n_{\bf p_1})(1+n_{\bf p_2})) \right]
\nonumber \\
&&\times (2\pi)^{d+1} \delta^{d}({\bf p_1+p_2-p_3-p})
 \delta(\Omega _{\bf p_1}+\Omega _{\bf p_2}-\Omega _{\bf p_3}-\Omega _{\bf p})
\; ,
\label{Bol}
\end{eqnarray}
where $\Omega_p=\sqrt{{\bf p}^2+\mu^2(t)}$ and 
the mass $\mu^2(t)$ is the self-consistent solution of 
\begin{eqnarray}
\mu(t)^2=m^2+ \frac{\lambda}{2}  \int \!\! \frac{d^{d}k}{(2\pi)^{d} }
\frac{n_{\bf k}(t)}{\sqrt{\mu(t)^2+{\bf k}^2}}  
\; .
\label{eq:BolMass}
\end{eqnarray}
In fact, this Boltzmann equation can be derived from 
the KB equations at the leading order in the gradient expansion
and with the Markov and quasi-particle approximations\cite{CH}.  
We remark here that in 1+1 dimensions 
the Boltzmann equation cannot lead to thermalization
because 
the particle momenta must be unchanged 
in each 2-to-2 collision in order to satisfy
the energy and momentum conservations.

\section{Entropy of the relativistic Kadanoff-Baym equations}
\label{sec:entropy}

In this section we derive the expression for the relativistic
entropy in terms of the two-point functions $G(x,y)$ 
for the $\lambda \phi^4$ theory, 
as an extension from the non-relativistic entropy current given
in \cite{IKV4} and \cite{Kita}.

We start with the Schwinger-Dyson equation (\ref{S-D}).
Multiplying $G$ from the right and left hand sides
of Eq.~(\ref{S-D}), respectively,
we obtain
\begin{eqnarray}
-\left[
\partial_x^2+m^2+ \frac{\lambda}{2}G^{aa}(x,x) 
\right] G^{ab}(x,y)
-i \int dz  \Sigma_{\rm nonl}^{ac}(x,z)c^{cd}G^{db}(z,y) 
=
ic^{ab}\delta(x-y)
\; ,
\label{SDR}
\\
-\left[
\partial_y^2+m^2+ \frac{\lambda}{2}G^{bb}(y,y)
\right]G^{ab}(x,y)
-i\int dz 
G(x,z)^{ac}c^{cd}\Sigma_{\rm nonl}^{db}(z,y)
=
ic^{ab}\delta(x-y)
\; ,
\label{SDL}
\end{eqnarray}
where $a$ and $b$ assign the branch 1 and 2 of the Schwinger-Keldysh
contour $\mathcal{C}$ and $c^{ab}={\rm diag}(1,-1)$.
We introduce the ``center-of-mass'' coordinate
$X=(x+y)/2$ and the relative coordinate $x-y$.
Then making the difference of these equations
(\ref{SDL}) and (\ref{SDR})
and performing the Fourier transform
with respect to the relative coordinate $x-y$,
we find
\begin{eqnarray}
&&
\left [
  2i p\cdot \frac{\partial}{\partial X}
  -\frac{i}{2}\cdot \frac{\lambda}{2} 
  \int \!\! \frac{d^{d+1}k} { (2\pi) ^{d+1}}
  \left (
    \frac{\partial G^{aa}(X,k)}{\partial X}
   +\frac{\partial G^{bb}(X,k)}{\partial X}
  \right)
  \cdot \frac{\partial}{\partial p} 
\right ] G^{ab}  
\nonumber \\
&=& i\int \! d(x-y) e^{ip\cdot (x-y)}
 \int \!dz (
     \Sigma^{ac}_{\rm nonl}(x,z)c^{cd}G^{db}(z,y)
     -G^{ac}(x,z)c^{cd}\Sigma_{\rm nonl}^{db}(z,y)
) \; ,
\label{dif}
\end{eqnarray}
where 
$p$ and $k$ are the momentum conjugate to $x-y$.
When we make the sum of them and perform the Fourier transform, we get
the expression 
\begin{eqnarray}
&&
\left[
   p^2-m^2 -\frac{\lambda}{4}
   \left(
      \int \!\! \frac{d^{d+1}k}{(2\pi)^{d+1}}
          ( G^{aa}(X,p)+G^{bb}(X,p) ) 
   \right)
 \right]G^{ab}(X,p)
\nonumber \\
&=&
ic^{ab} + 
\frac{i}{2}\int \! d(x-y) e^{-ip\cdot (x-y)} 
\int\! dz(\Sigma_{\rm nonl}^{ac}(x,z)c^{cd}G^{db}(z,y)
         +G^{ac}(x,z)c^{cd}\Sigma^{db}_{\rm nonl}(z,y)).
\label{ave}
\end{eqnarray}
Starting the evolution at $x^0=y^0=0$,
we have $G(x,y)$ only in a finite region of $x^0$ and $y^0$.
It is therefore important to note that 
the interval of $x^0 - y^0$ is inevitably
limited within $\pm X^0$ in the Fourier transformation.

The gradient expansion with respect to the center-of-mass coordinate
$X$ is adequate when the $X$-dependence of the system is smooth enough
(See for example \cite{{BM},{IKV4}}).
We keep just the first order terms in the gradient
expansion of Green's functions and the self energies here. 
For the expansion of the right hand side of
Eqs.~(\ref{dif}) and (\ref{ave}), we use the formula
for two point functions $K(x,y)$ and $L (x,y)$:
\begin{eqnarray}
\int d(x-y) e^{i p\cdot (x-y)} \int dz K(x,z) L(z,y) 
&= &
\tilde K(X,p) \tilde L(X,p)
\nonumber \\
&&
+ \frac{i}{2} 
\left( \frac{\partial \tilde K}{\partial p^\mu} 
       \frac{\partial \tilde L}{\partial X_\mu}
    -  \frac{\partial \tilde K}{\partial X^\mu} 
       \frac{\partial \tilde L}{\partial p_\mu} 
\right)
+ O\left(\frac{\partial^2}{\partial X^2}\right)
\; ,
\label{gra}
\end{eqnarray}
where $\tilde K(X,p)$ and $\tilde L (X,p)$ are the Fourier-transforms
in $x-y$.
We remark here 
the  scale separation between $X^0$ and $x^0-y^0$.
We implicitly assume that 
the time dependence on the former is smooth and mild while
the time correlation in the latter is much shorter.
At the very early time of the evolution,
this separation cannot be expected and the gradient expansion 
should be invalid.

The derivation of the entropy current is most simplified in terms of
the retarded propagator $G_R=i(G^{11}-G^{12})$. 
To the first order in the gradient expansion,
Eqs.~(\ref{dif}) and (\ref{ave}) reduce to the equations for $G_R$:
\begin{eqnarray}
\left[ 
   \frac{\partial \left (M -\tfrac{1}{2}\Sigma_\rho\right)}
        {\partial p^\mu}
   \frac{\partial }{\partial X_\mu} 
 - \frac{\partial \left( M-\tfrac{1}{2}\Sigma_\rho \right)}
        {\partial X^\mu} 
   \frac{\partial}{\partial p_\mu}
\right]
 G_R(X,p) &=& 0
\; ,
\label{grr1}
\\
\left (M-\tfrac{1}{2}\Sigma_\rho  \right )G_R(X,p) &=& -1
\; ,
\label{grr2}
\end{eqnarray}
where $M$ denotes\footnote{Please don't confuse this $M$ 
with a mass function.
Both $M$ and $\Sigma$ have mass-dimension 2.} 
\begin{eqnarray}
M=p^2-m^2 -{\rm Re} \Sigma_{R},   
\quad
\Sigma_{R}=\Sigma^{11}-\Sigma^{12}
\; .
\end{eqnarray}
In deriving Eqs.~(\ref{grr1}) and (\ref{grr2}) 
we have used the well-known relations
$\Sigma^{11}+\Sigma^{22}=\Sigma^{12}+\Sigma^{21}$,
$G^{11}+G^{22}=G^{12}+G^{21}$
and
$2i{\rm Im}\Sigma_R=\Sigma_\rho$.  
The formal solution of the above simultaneous equations (\ref{grr1})
and (\ref{grr2}) is written as \cite{BM} 
\begin{eqnarray}
G_R(X,p) = \frac{-1}{M-\tfrac{1}{2}\Sigma_\rho } 
\; .
\end{eqnarray}
One should note here that $M$ ($\Sigma_\rho$)
is real (imaginary).
Therefore, the real and imaginary parts of the retarded propagator
$G_R$ are given as
\begin{eqnarray}
{\rm Re} G_R(X,p) &=& -\frac{M}{M^2 - \frac{1}{4} \Sigma_\rho ^2 }  
\; ,
\label{GR}
\\
\rho(X,p) =2i{\rm Im}G_R(X,p) &=&
- \frac{\Sigma_\rho}{ M^2- \frac{1}{4}\Sigma_\rho ^2 }
\; .
\label{rfg}
\end{eqnarray}
We see that the spectral function $\rho(X,p)$ 
has the Breit-Wigner form
(\ref{eq:rfree}) in the first order approximation of  the gradient
expansion.

Now we are ready for writing down the entropy current.
The ``derivation'' goes somewhat in a heuristic way.
We make the difference of 
Eq.~(\ref{dif}) for $(a,b)=(1,2)$ multiplied by $\ln (iG^{12}/\rho)$
and 
Eq. (\ref{dif}) for $(a,b)=(2,1)$ multiplied by $\ln (iG^{21}/\rho)$.
Then we integrate the resultant expression 
over $d^{d+1} p/ (2\pi)^{d+1}$ to arrive
at the following equation:
\begin{eqnarray}
\partial _\mu s^{\mu} 
=
\frac{1}{2} \int \!\! 
\frac{d^{d+1}p}{(2\pi)^{d+1}} \ln \frac{G^{12}}{G^{21}} C(X,p) .
\label{entropy}
\end{eqnarray} 
Here the term $C$,
\begin{eqnarray}
C(X,p)=i \left (
       \Sigma_\rho(X,p)\; F(X,p) -\Sigma_F(X,p)\;  \rho(X,p)
        \right ) ,
\end{eqnarray}
may be identified as the collision term in the Boltzmann limit.
With Eq.~(\ref{entropy}), we define the entropy current $s^\mu (X)$ as
\begin{eqnarray}
s^\mu 
&=&
 \int \!\! \frac{d^{d+1}p}{(2\pi)^{d+1} } 
\left [
\left(
   p^\mu -\frac{1}{2}\frac{\partial {\rm Re} \Sigma_R}{\partial p_\mu} 
\right)
\left ( 
   -G^{12}\ln \frac{iG^{12}}{\rho}+G^{21} \ln \frac{i G^{21}}{\rho}
\right) 
\right .
\nonumber \\
&&
-\frac{1}{2} {\rm Re} G_R \
\left .
\left(
   -\frac{\partial}{\partial p_\mu} 
    \left(  \frac{\Sigma_\rho}{i} \frac{iG^{12}}{\rho}  \right)
\ln \frac{iG^{12}}{\rho} 
   + \frac{\partial}{\partial p_\mu} 
     \left (  \frac{\Sigma_\rho}{i} \frac{iG^{21}}{\rho}
     \right )
\ln \frac{iG^{21}}{\rho}  
\right ) 
\right  ]
\, ,
\end{eqnarray}
where 
we have used the relations $i(\Sigma^{11}-\Sigma^{22})=2{\rm Re}\Sigma_R$
and  $i(G^{11}-G^{22})=2{\rm Re} G_R$.
We have also applied the approximations
$\Sigma^{12} \simeq \Sigma_\rho \frac{G^{12}}{\rho}$
and $\Sigma^{21} \simeq \Sigma_\rho \frac{G^{21}}{\rho}$
in the first order gradient expansion
\cite{IKV4},\cite{Cassing}.

When we write the two-point functions
in the form of the Kadanoff-Baym Ansatz
$G^{12}=-i\rho f$ and $G^{21}=-i\rho (1+f)$ with
a real function $f$,
the above expression for the entropy current becomes
\begin{eqnarray}
s^\mu
&=&
\int \!\! \frac{d^{d+1}p}{(2\pi)^{d+1} } 
\left \{
\frac{\rho}{i} 
\left( 
    p^\mu -\frac{\partial {\rm Re} \Sigma_R}{\partial p_\mu}
\right) 
\left( -f \ln f+(1+f)\ln (1+f) \right)
\right .
\nonumber \\
&&
\left .
-\frac{1}{2} {\rm Re} G_R 
\left [ 
   -\frac{\partial}{\partial p_\mu} 
    \left( \frac{\Sigma_\rho}{i} f\right) 
\ln f \;   
 +\frac{\partial}{\partial p_\mu } 
    \left( \frac{\Sigma_\rho}{i} (1+f) \right)
 \ln (1+f) \; 
    \right ]
\right \}
\;  .
\end{eqnarray}
After integration by parts over $p^\mu$ in the second line, we obtain
a simple expression:
\begin{eqnarray}
s^\mu = \int \!\! \frac{d^{d+1}p}{(2\pi)^{d+1} } 
 \left[ \frac{\rho}{i} \left( p^\mu-\frac{1}{2} \frac{\partial {\rm Re} \Sigma_R}{ \partial p_\mu} \right) 
+ \frac{ \Sigma_\rho}{i} \frac{1}{2}\frac { \partial {\rm Re} G_R}{\partial p_\mu} \right] \sigma
\; ,
\label{eq:s3}
\end{eqnarray}
where we introduced the notation
\begin{eqnarray}
\sigma(X,p) = -f \ln f +(1+f) \ln (1+f) \; .
\end{eqnarray}
One must distinguish
this ``occupation number'' function $f$ in the Kadanoff-Baym Ansatz 
from the distribution function $n_{\bf p}$ defined in (\ref{eq:np}).

Substituting the solution (\ref{GR}) for $G_R$,  we can write
the entropy current more explicitly as
\begin{eqnarray}
s^\mu 
&=&
\int \!\! \frac{d^{d+1}p}{(2\pi)^{d+1} } 
\left [ 
  \frac{\rho}{i} 
  \left(
     1+ \frac{ M^2-\frac{ \Sigma_\rho^2}{4} -2  M^2}
             {M^2 -\frac{\Sigma_\rho^2}{4} } 
  \right) 
  \left( 
     p^\mu -\frac{1}{2} 
     \frac{\partial {\rm Re} \Sigma _R}{\partial p_\mu }
  \right) 
+ 
 \frac{\rho}{4i}
 \frac {M \Sigma_\rho \frac{\partial \Sigma_\rho}{\partial p_\mu}}
       { M^2 - \frac{\Sigma _\rho^2}{4} } 
\right] \sigma
\; .
\end{eqnarray}
This expression further simplifies with use of (\ref{rfg})
to%
\footnote{
The tadpole part should be the renormalized one in this expression in 1+1
dimensions.}
\begin{eqnarray}
s^\mu = \int \!\! \frac{d^{d+1}p}{(2\pi)^{d+1} }  \frac{ \rho^2 \Sigma_\rho}{ 2i} 
\left [ \left(p^\mu -\frac{1}{2} \frac{\partial {\rm Re} \Sigma _R} {\partial p_\mu }  \right)
-\frac{1}{2} \frac {M}{\Sigma _\rho} \frac{\partial \Sigma _\rho}{\partial p_\mu} \right] \sigma
\; .
\label{eq:entg}
\end{eqnarray}
This is one of the main results of this work.
This expression of the entropy current is a natural extension
to the relativistic case.
The only difference between non-relativistic \cite{{IKV4},{Kita}} and our
relativistic case is the factor $\frac{1}{2}$ in front of 
the momentum derivative of the self-energy.
We remark here that there is a discussion about the memory correction
terms to the kinetic entropy in the non-relativistic case
in Refs.~\cite{IKV4} and \cite{Kita} when we deal with the skeleton
diagrams $\Sigma_{\rm nonl}$ beyond the NLO in $\lambda$.

In the quasi-particle limit,
$\Sigma_{\rm nonl} \rightarrow 0$,
we know that
$G^{12}=-i\rho f=
2 \pi\delta ((p^0)^{2}- \Omega_{\bf p}^2) (\theta(-p^0)+n_{\bf p})$
and 
$G^{21}= -i \rho (1+f) =
2\pi \delta ((p^0)^{2}- \Omega_{\bf p}^2)(\theta(p^0)+n_{\bf p}) $.
In this limit 
the expression of the entropy current for $\mu=0$
reduces to the well-known form of the entropy density for bosons 
\begin{eqnarray}
s^0 = \int \!\! \frac{d^{d}p}{(2\pi)^{d} } 
\left[ -n_{\bf p}\ln n_{\bf p} + (1+n_{\bf p}) \ln (1+n_{\bf p}) \right]
\; ,
\label{eq:entf}
\end{eqnarray}
as it should be.
In general cases, however,
the spectral function $\rho(X,p)$ is defined
as the Fourier transform of Eq.~(\ref{eq:rho}) in $x-y$,
and the occupation number function $f(X,p)$ is then obtained 
with $G^{12}(X,p)=-i \rho(X,p) f(X,p)$.
Although $p^0 \rho(X,p) \ge 0$ in equilibrium,
we have only a finite support in $x^0-y^0$ in the initial value problem
and the resultant Fourier transform $ p^0 \rho(X,p)$ may oscillate
in $p^0$, as shown in the next section.
Accordingly the function $f(X,p)$ can become negative,
which brings a difficulty in evaluating the entropy density
$s^0$ obtained at the leading order of the gradient expansion.


Finally we show the fact 
that this entropy current obtained
in the NLO in $\lambda$ 
formally satisfies the H-theorem.
Namely,
the RHS of Eq.~(\ref{entropy}) is positive semi-definite.
This can be verified by substituting the expressions
for $\Sigma_F$ (\ref{sigf}) and $\Sigma_\rho$ (\ref{sigr})
into the RHS of (\ref{entropy}).
As a result we obtain the relation
\begin{eqnarray}
 \partial_\mu s^\mu (X) 
&=& \int \!\! 
\frac{d^{d+1}p}{(2\pi)^{d+1}}
\frac{1}{2} \ln \frac{G^{12}}{G^{21}} C
\nonumber \\
&=&
\frac{1}{8}\cdot \frac{\lambda^2}{3!} 
\int \!\! \frac{d^{d+1}p}{(2\pi)^{d+1}}
\frac{d^{d+1}q}{(2\pi)^{d+1}}
\frac{d^{d+1}k}{(2\pi)^{d+1}}
\frac{d^{d+1}r}{(2\pi)^{d+1}}
(2\pi)^{d+1}(p+q-k-r) 
\nonumber \\
&\times& \Big[ G^{12}(p,X)G^{12}(q,X) G^{21}(k,X) G^{21}(r,X) 
- G^{21}(p,X)G^{21}(q,X) G^{12}(k,X) G^{12}(r,X)  \Big]
\nonumber \\
&\times& \ln \frac {G^{12}(p,X)G^{12}(q,X) G^{21}(k,X) G^{21}(r,X) 
}{G^{21}(p,X)G^{21}(q,X) G^{12}(k,X) G^{12}(r,X)  } \geq 0
\; ,
\label{htheorem}
\end{eqnarray}
where we have used $G^{12}(-k,X)=G^{21}(k,X)$.
The last inequality holds 
since $(x-y)\ln \frac{x}{y} \geq 0$.
Thus we proved that the H-theorem is fulfilled
in the NLO
in $\lambda\phi^4$ theory. 
However, at higher orders in 
the coupling constant $\lambda$,
the definition of the entropy current
and the proof of the H-theorem
are open problems.

\section{Numerical simulation}
\label{sec:numerics}

\begin{figure}[t]
\begin{minipage}{0.45\textwidth}
\begin{center}
\includegraphics[ height=\textwidth, angle=270 ]{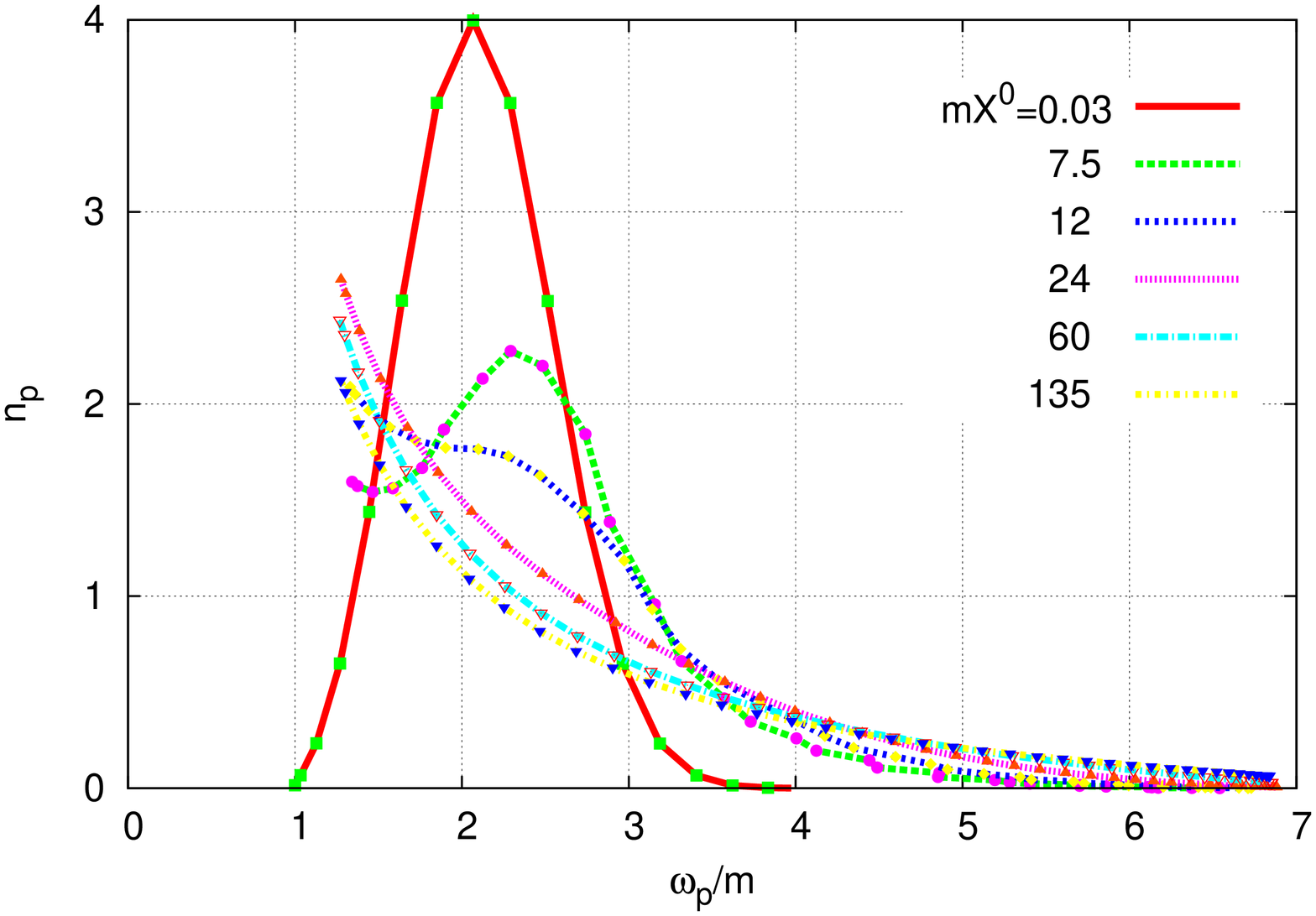}
\caption{Evolution of the distribution function 
$n_{\bf p} (\tilde \omega_{\bf p} / m)$ from the tsunami initial
 condition ($\lambda/m^2=4$).}
\label{fig:num}
\end{center}
\end{minipage}
\hfil
\begin{minipage}{0.45\textwidth}
\begin{center}
\includegraphics[ height=\textwidth, angle=270 ]{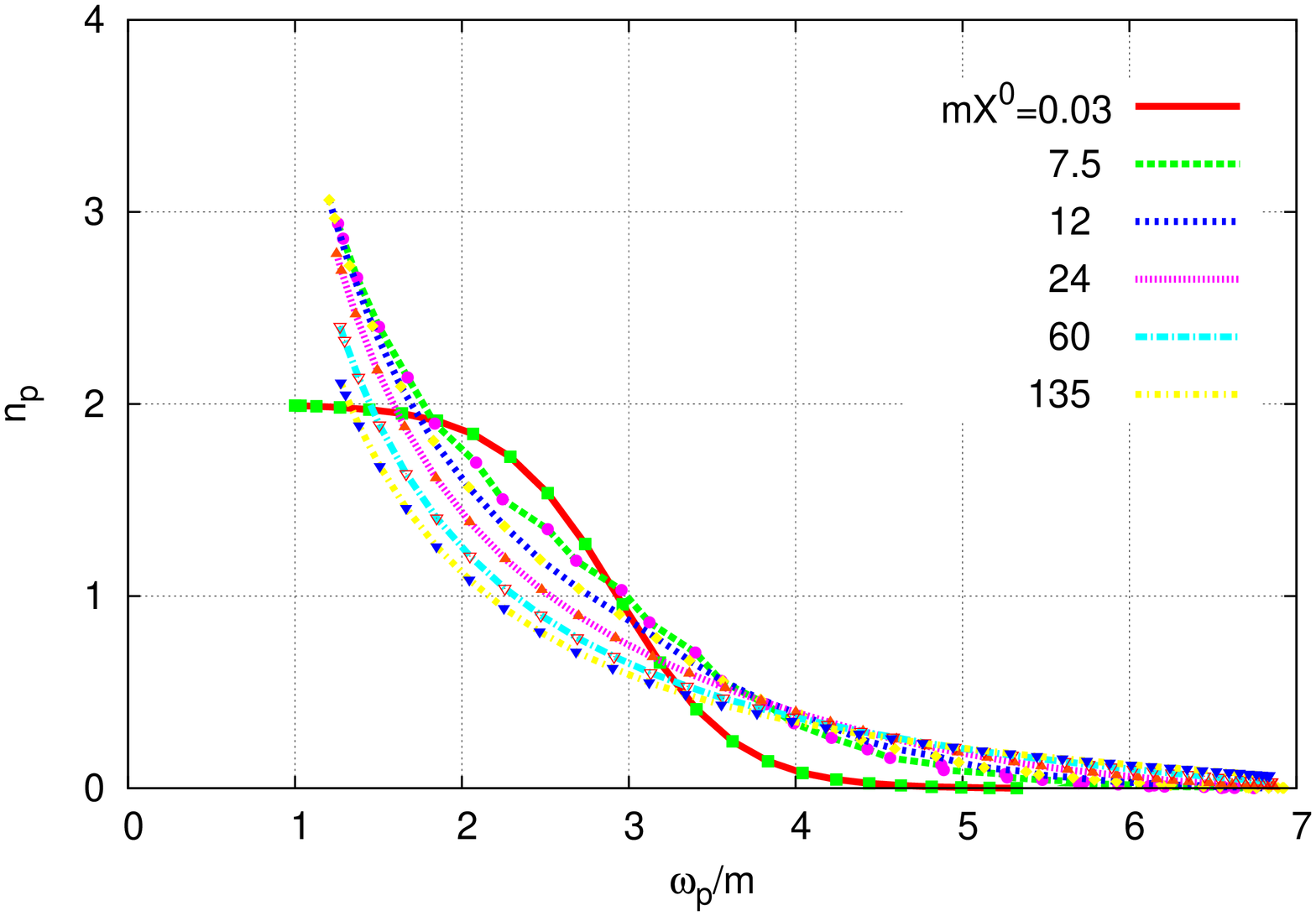}
\caption{Evolution of the distribution function
$n_{\bf p}(\tilde \omega_{\bf p} / m)$ from the WS initial 
condition ($\lambda/m^2=4$).}
\label{fig:num2}
\end{center}
\end{minipage}
\end{figure}

\begin{figure}[t]
\begin{center}
\includegraphics[height=0.45\textwidth, angle=270 ]{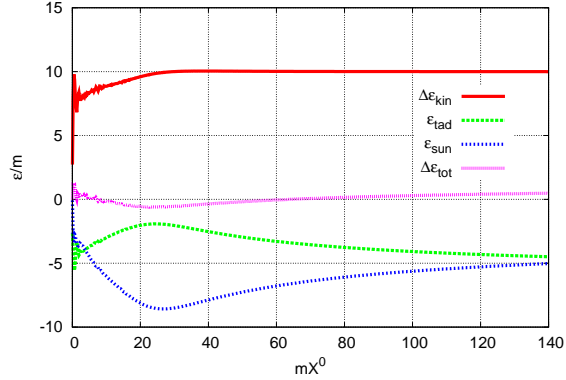}
\caption{Evolution of the energy content in units of $m$
from the tsunami initial condition with $\lambda/m^2=4$;
Kinetic $\epsilon_{\rm kin}/m$ (solid), 
tadpole $\epsilon_{\rm tad}/m$ (dashed), 
sunset $\epsilon_{\rm sun}/m$ (dotted)
and the total energy $\epsilon_{\rm tot}/m$ (bold solid).
See the text for the details.}
\label{fig:Ene}
\end{center}
\hfil
\end{figure}

In this section we show numerical results of time evolutions
of the KB equations in the  $\lambda \phi^4$ theory in 1+1 dimensions. 
We discretize the space $L= 2 N  a_s$ into $2N$ grid points
$x_n= n a_s$ ($n=-N, -N+1, \cdots,  N-1, N$) with $a_s$ the lattice
spacing and apply the periodic boundary condition.
Accordingly the momentum has discrete value as $p_n= \frac{2 \pi n}{L}$.
The space derivative $-\partial_x^2$ is replaced 
with $\hat p^2=\frac{4}{a_s^2} \sin^2\left(\frac{a_s p_n}{2} \right)$,
which removes much of the lattice artifacts\cite{IMGM}.
We set $N=40$, which is sufficient to study the momentum dependence.
We also performed the simulation
with $N=80$ and found no appreciable differences in the numerical results.
We solve the evolution with the time step, $a_t/a_s=0.1$.

We set the mass $m a_s=0.3$ and varied 
the coupling $\lambda/m^2$=4, 2 and 1.
We prepared the two different types of the
initial conditions, ``tsunami'' distribution and the Woods-Saxon (WS)
distribution, respectively,
\begin{eqnarray}
n^T_{\bf p}=
\frac{1}{{\cal N}_T}
\exp\left[ - \frac{(|p_x|-{\bf p}_T)^2}{2 \sigma^2}\right]
\label{eq:icT}
\end{eqnarray}
with $\sigma^2/m^2=4.4\times \left(\frac{2\pi}{mL} \right)^2$, $p_T= 7\cdot 2 \pi /L$
and $\mathcal{N}_T=0.25$,
and
\begin{eqnarray}
n^{WS}_{\bf p}
=\frac{1}{ {\cal N}_{_{WS}} }\;
\frac{1}{e^{(\sqrt{|{\bf p}|^2+m^2}-p_{_{WS}})/\kappa}+1}
\label{eq:icF}
\end{eqnarray}
with $p_{_{WS}}/m=2.936$, $\kappa/m=0.35$
and $\mathcal{N}_{_{WS}}=0.5$.
The ``tsunami'' initial condition has two peaks at $\pm {\bf p}_T$ (``tsunami'') 
and may be regarded as a toy model of the nuclear collisions.
The WS initial condition is used to check the sensitivity of the
evolution to the initial condition.
These parameters in the WS case are tuned
so that both the initial conditions give the same energy for
$\lambda/m^2=4$. 
Later in this section, we change the coupling constant $\lambda$ with
other parameters fixed, in order to see how the evolution depends on
the coupling strength.
We monitor the energy conservation in each time step in order to
estimate the numerical accuracy of our simulation.

In Fig.~\ref{fig:num} we show the time evolution of the number
distribution $n(\tilde \omega_{\bf p})$ defined in 
Eqs.~(\ref{eq:np}) and (\ref{eq:op}) with the tsunami
initial condition (\ref{eq:icT}).
From this figure we confirm that our simulation reproduces the
results of Ref.~\cite{AB}. 
The peak of Gaussian distribution disappears rapidly and
the values at high and low edge regions grow up with time. 
As a result, the particle number distribution approaches the Bose distribution
function  $n_{\bf p}= 1/(e^{ \epsilon_{\bf p}/T}-1)$, 
with temperature ${T}/{m}\sim 2.5$ and zero chemical potential.
Similarly in Fig. \ref{fig:num2}
we show the time evolution of number distribution 
$n(\tilde \omega_{\bf p})$
with the WS initial condition (\ref{eq:icF}). 
We see that $n(\tilde \omega_{\bf p})$
converges to the same thermal distribution as
the one in the case of ``tsunami'' initial condition.

Next we study the energy content of the system. 
The explicit expression of the energy is given
in Appendix \ref{sec:ene}.
We plot
the kinetic (\ref{eq:ekin}), 
tadpole (\ref{eq:etad}), sunset (\ref{eq:esun}) and total
energy (\ref{eq:ecomp}) as a function of time $m X^0$ in Fig.~\ref{fig:Ene}.
As for the total and kinetic energies we plot their differences
measured from the initial value of the total energy
 $\epsilon_{\rm tot}/m \sim 260$ in our discretization.
We find that the growth of the kinetic energy is canceled by the
tadpole and the sunset energy to have a constant total energy. 
The energy is conserved within 0.5 percent in Fig.~\ref{fig:Ene}.

\begin{figure}[t]
\begin{minipage}{0.45\textwidth}
\begin{center}
\includegraphics[height=\textwidth, angle=270 ]{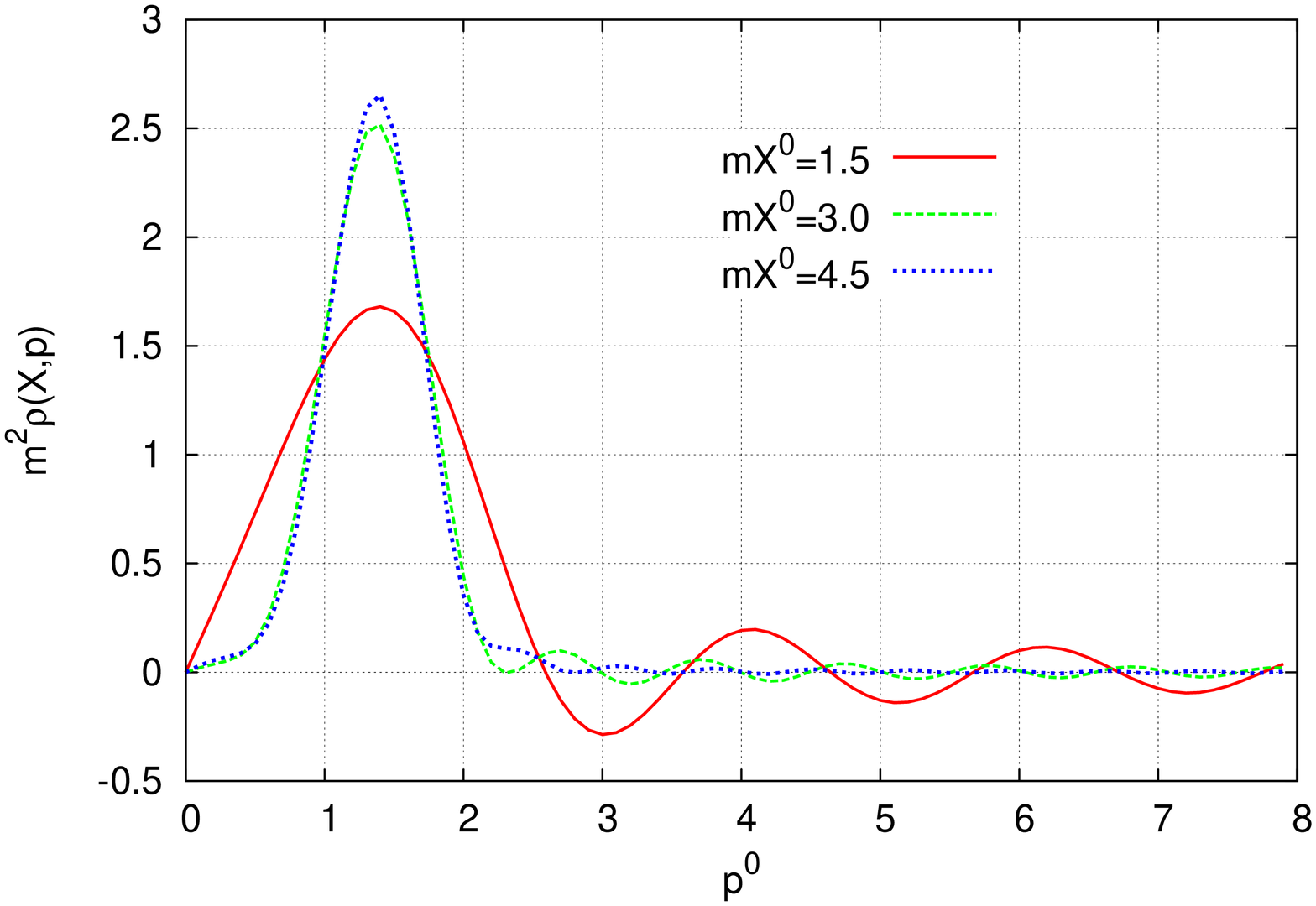}
\caption{Spectral function $\rho(X,p^0,p_x)$  with
$p_x=0$ at $mX^0 =$1.5, 3 and 4.5.}
\label{fig:sp1}
\end{center}
\end{minipage}
\hfil
\begin{minipage}{0.45\textwidth}
\begin{center}
\includegraphics[height=\textwidth, angle=270 ]{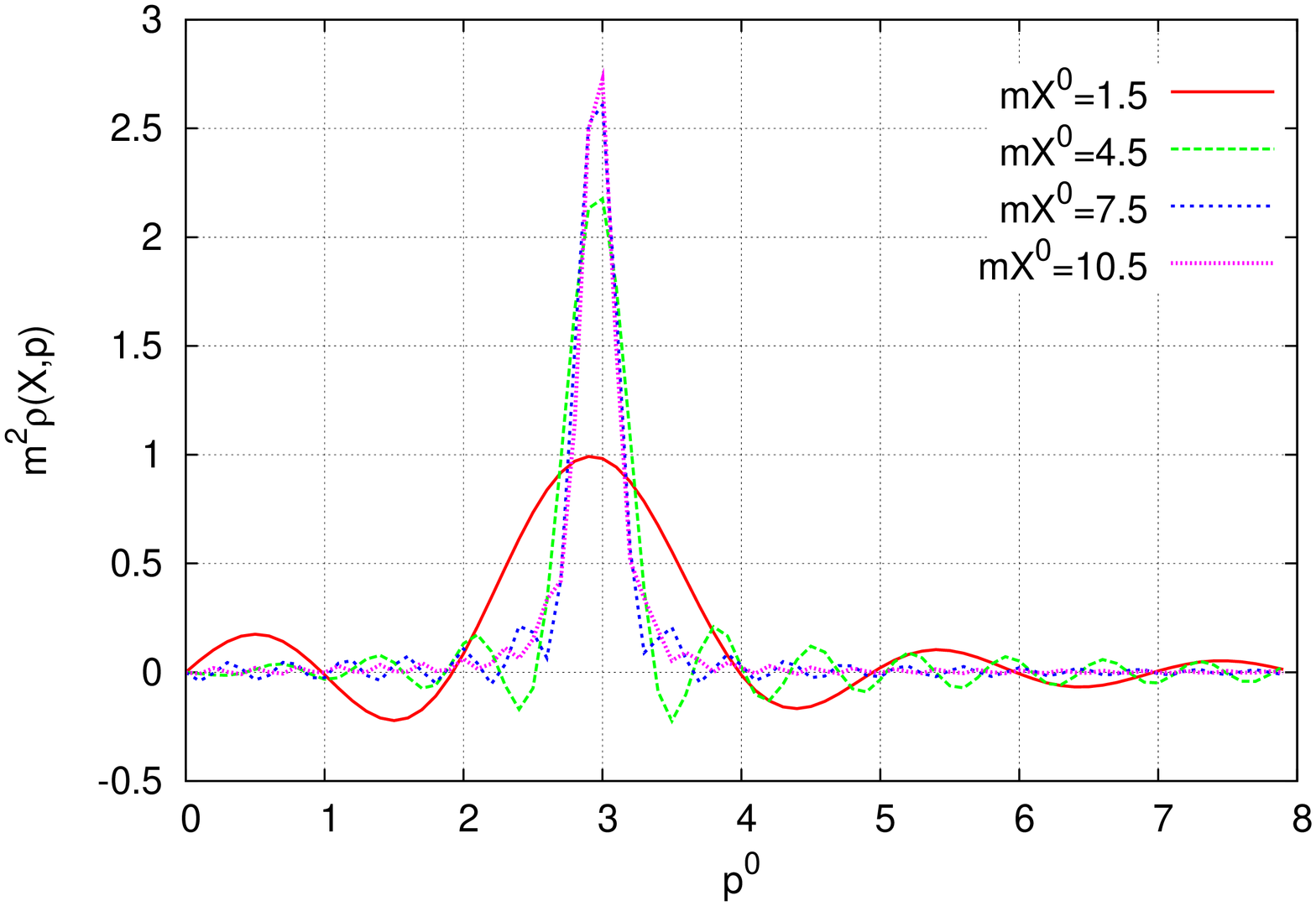}
\caption{ Spectral function $\rho(X,p^0,p_x)$ with
$p_x=20\pi /L$ at $mX^0 =1.5, \cdots, 10.5$.}
\label{fig:sp2}
\end{center}
\end{minipage}
\end{figure}

\begin{figure}[t]
\begin{center}
\includegraphics[height=0.45\textwidth, angle=270 ]{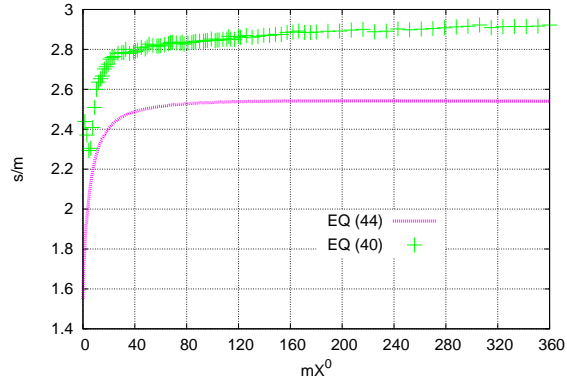}
\caption{ Time evolution of kinetic entropy (\ref{eq:s3}) denoted in
$+$ and its quasi-particle approximation (\ref{eq:entf}) shown in a
curve for the ``tsunami'' initial condition with coupling $\lambda/m^2=4$. }
\label{fig:ent}
\end{center}
\end{figure}

\begin{figure}[htbp]
\begin{minipage}{0.45\textwidth}
\begin{center}
\includegraphics[height=\textwidth, angle=270 ]{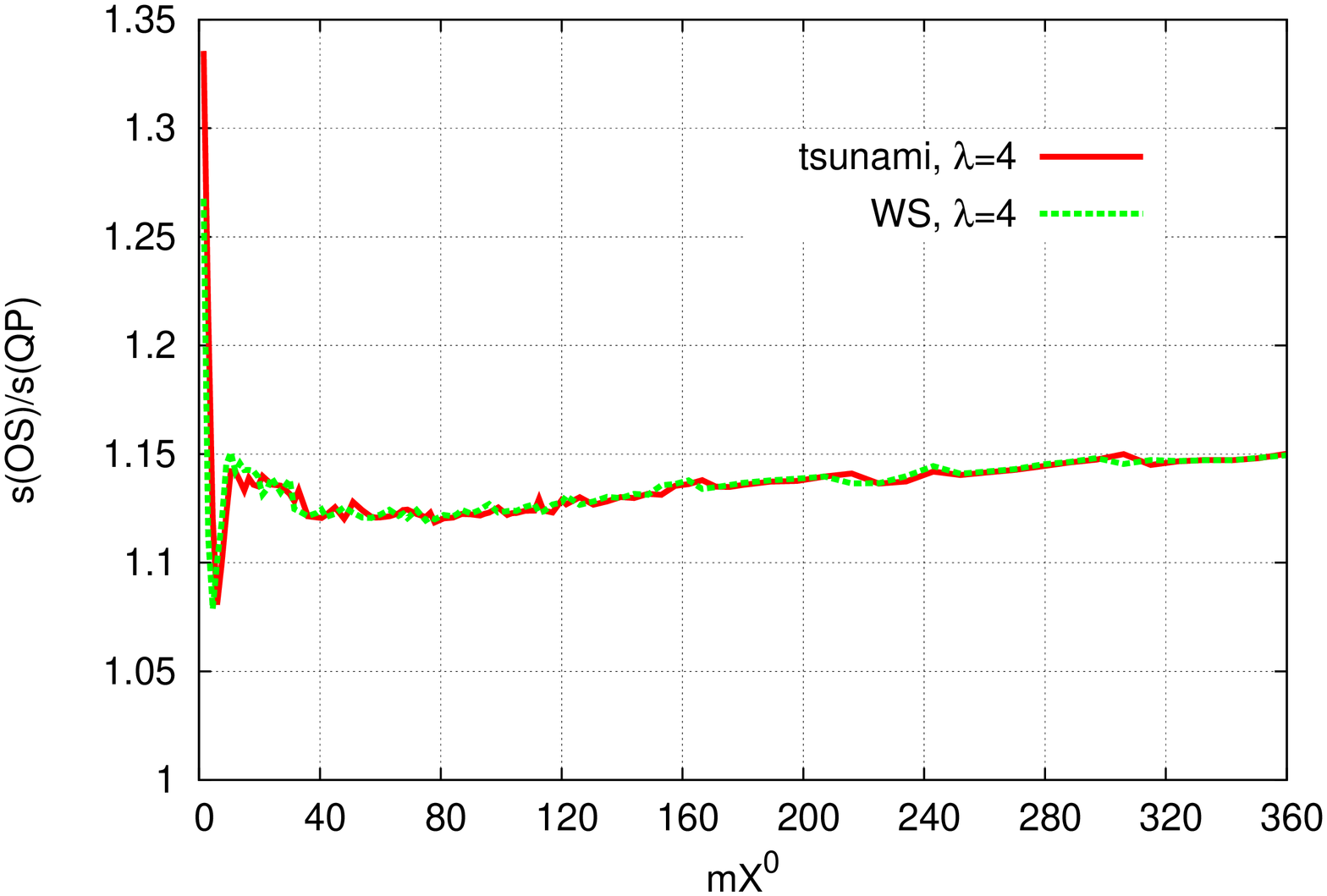}
\caption{The ratio between kinetic entropy (40) and its quasiparticle approximation (44) for 
the tsunami and WS initial conditions with $\lambda=4$.}
\label{fig:ENTR}
\end{center}
\end{minipage}
\hfil
\begin{minipage}{0.45\textwidth}
\begin{center}
\includegraphics[height=\textwidth, angle=270 ]{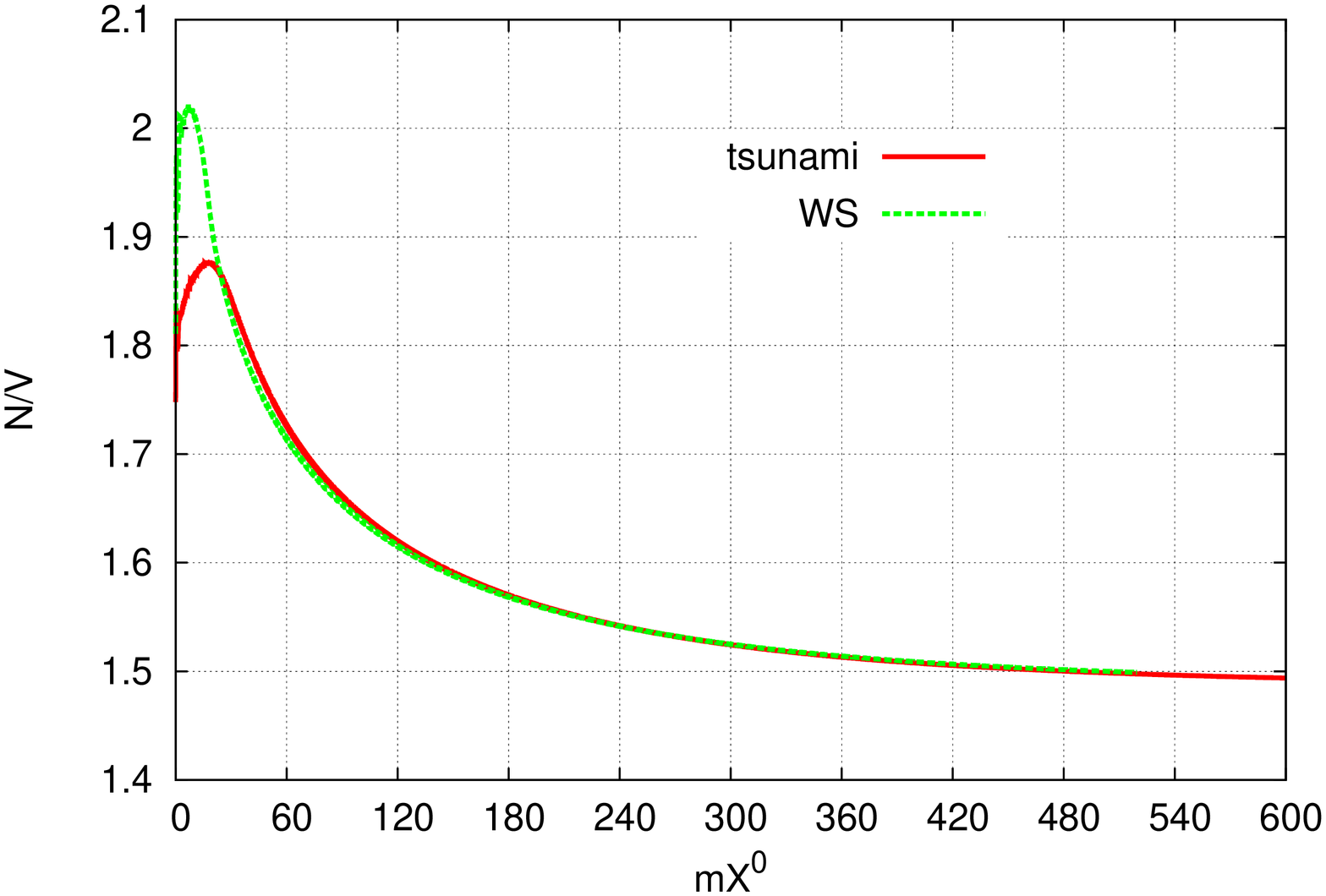}
\caption{Time evolution of total number density for the tsunami and WS
initial conditions with $\lambda=4$.}
\label{fig:ND}
\end{center}
\end{minipage}
\end{figure}

\subsection{Kinetic entropy}

Let us study the kinetic entropy (\ref{eq:s3}) derived with the gradient
expansion of the KB equation. 
To this end we first examine the shape of the spectral function
$\rho(X,p)$, which appears in the expression (\ref{eq:s3}) 
and is needed to compute the occupation number function $f(X,p)$ in
$\sigma$. The spectral function $\rho(X,p)$ itself is physically important.
In Figs.~\ref{fig:sp1} and \ref{fig:sp2}, 
we show $\rho(X,p)$ for $p_x=2\pi n/L$ with $n=0$, and 10, respectively,
at several values of time $X^0$, with the ``tsunami'' initial condition
(\ref{eq:icT}).
We clearly see peak structures near $p^0/m \sim \sqrt{m^2+p_x^2}/m$ at later
times in both figures.

At early times, however, the spectral function $\rho(X,p)$ shows
oscillatory behavior. This can be understood as the uncertainty
relation between the energy and the time.
In the observation within a finite time interval $|x^0 - y^0|<X^0$,
one can resolve the $p^0$ dependence of 
$\rho(X,p)$ on the scale larger than $1/X^0$, because
we have an oscillating factor due to
$\int_{-X_0}^{X^0} dt \; \exp(-ip^0 t)=2\sin(p^0 X^0)/p^0$. 
We numerically confirmed that the oscillation frequency
is indeed proportional to $X^0$.
This means that any finer structure of $\rho(X,p)$ 
than a scale $1/T$ is resolved only after the evolution time of $X^0>T$.
The sharper the peak structure is, the longer time it needs to be 
resolved, which is seen by comparing the cases with $p_x=0$ and $20\pi /L$
as shown in Figs.~\ref{fig:sp1} and \ref{fig:sp2}, respectively. 

The applicable range of the kinetic entropy \cite{IKV4,Kita}
has not yet been examined so far 
with using the numerical solution of the KB
equation before this study.
Given the oscillating $\rho(X,p)$ near the initial time, 
we also have an oscillation for the occupation number function
$f(X,p)$ accordingly. We thus encounter a problem in evaluating 
$\sigma$ as it contains the logarithm of $f(X,p)$.
Here we come to recognize that
the form of the kinetic entropy (\ref{eq:s3}) obtained in the 
leading-order gradient expansion
cannot be applied in the early stage of the initial value problem,
although the gradient expansion itself becomes unjustifiable
there.

In Fig.~\ref{fig:ent} we try evaluating the kinetic entropy (\ref{eq:s3}) 
as a function of time $X^0$. 
Crude as it is, we simply neglect the contributions from the region of 
$p^0$  where $\rho(X,p)$ has negative values as an exploratory estimate.
We obtain the entropy that decreases until $mX^0\sim 10$. This peculiar
behavior is presumably stemming from omitting the negative $\rho(X,p)$
contribution, which seems likely to overestimate the entropy. 
In the later stage, say $m X^0>20$, the kinetic entropy increases
monotonically as time proceeds, which is expected from the fact that
the kinetic entropy (\ref{eq:s3}) satisfies the H-theorem. 

\begin{figure}[t]
\begin{minipage}{0.45\textwidth}
\begin{center}
\includegraphics[height=\textwidth, angle=270  ]{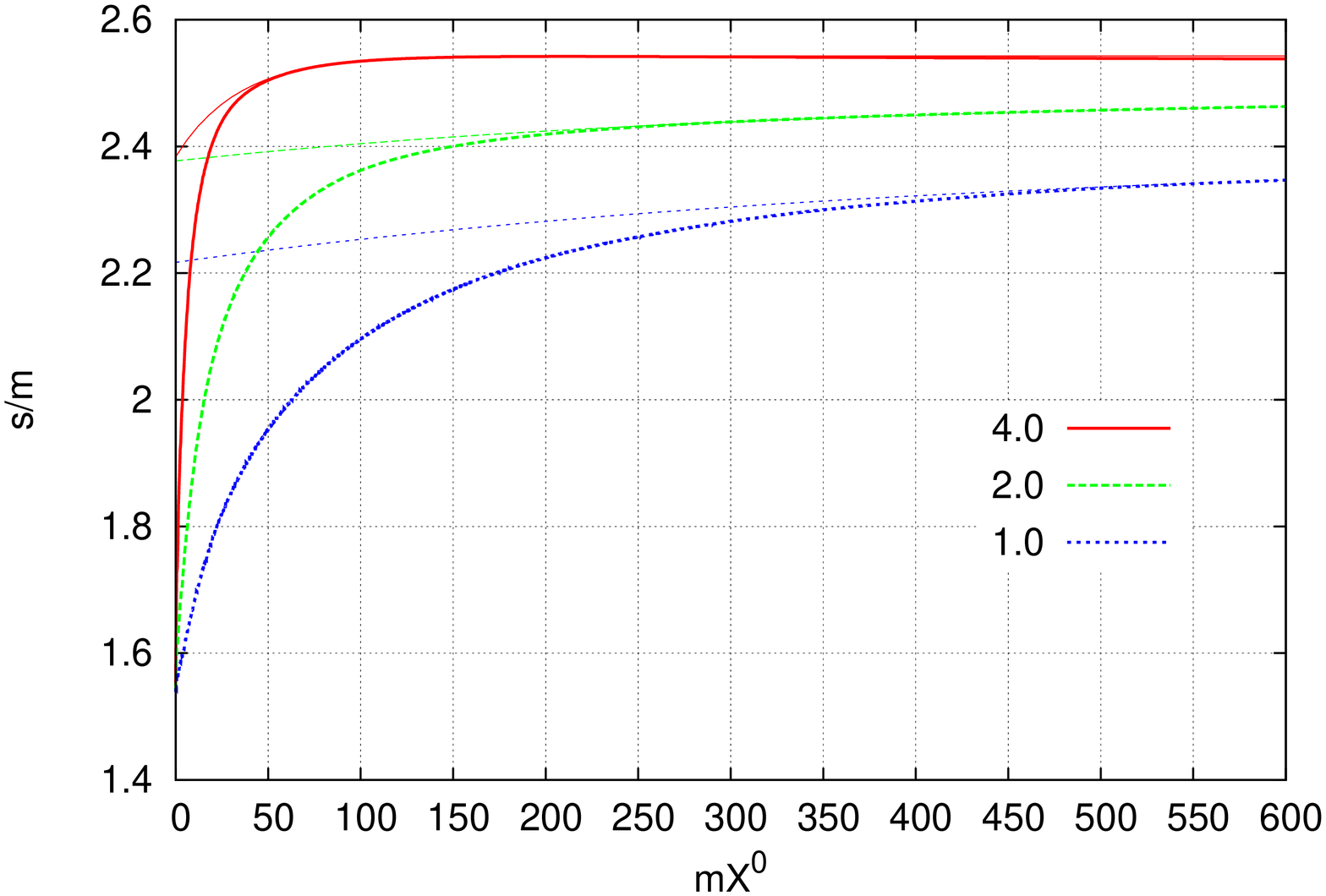}
\caption{Entropy density $s^0/m$ in the quasi-particle approximation 
for the "tsunami" initial condition
with coupling $\lambda/m^2=$4, 2 and 1. 
The exponential fit with (\ref{eq:fit})
is denoted by a thin line in each case.}
\label{fig:ent1}
\end{center}
\end{minipage}
\hfil
\begin{minipage}{0.45\hsize}
\begin{center}
\includegraphics[height=\textwidth, angle=270 ]{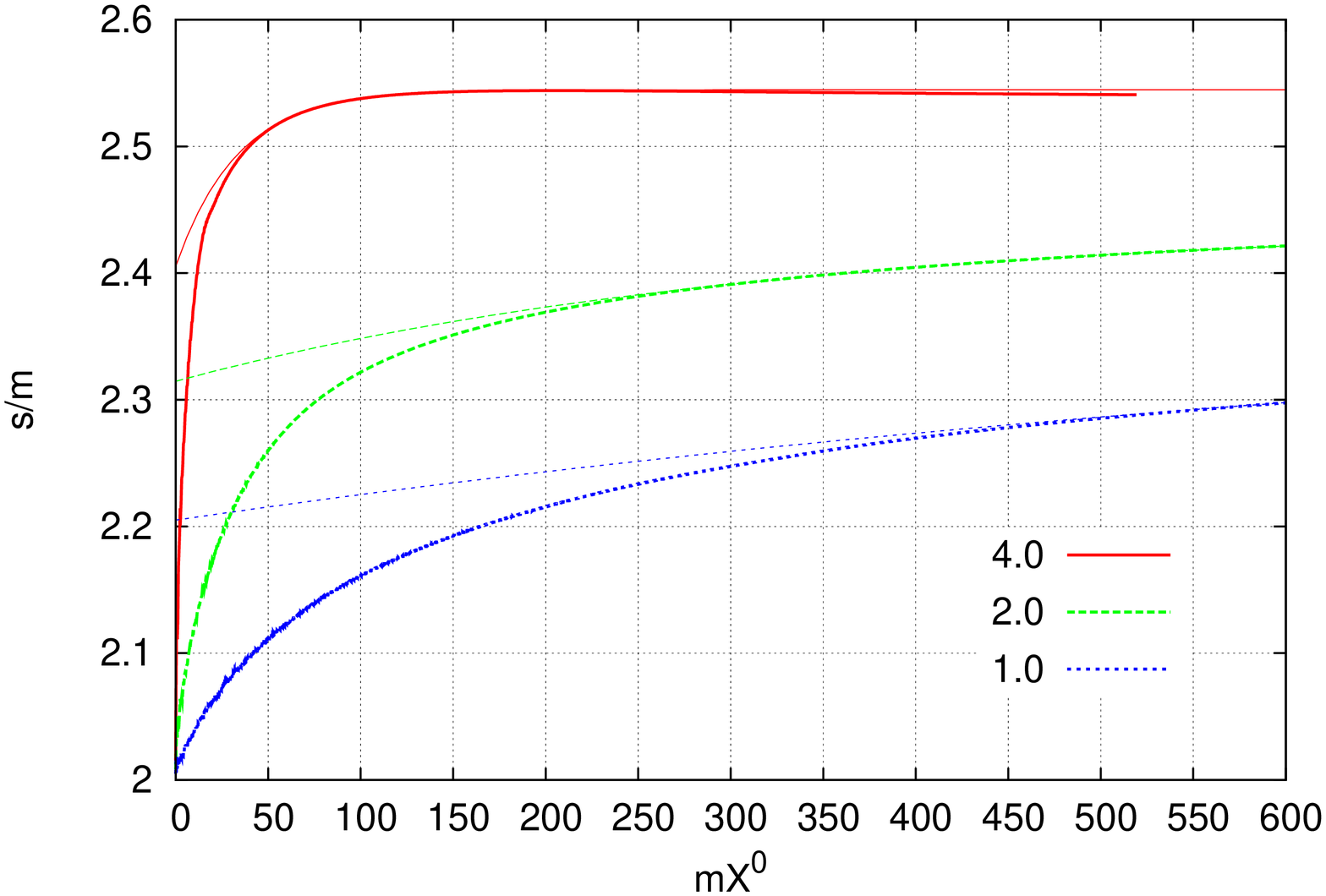}
\caption{Entropy density $s^0/m$ in the quasi-particle approximation
for the WS initial condition with coupling $\lambda/m^2=$4, 2 and 1.
The exponential fit with (\ref{eq:fit})
is denoted by a thin line in each case.}
\label{fig:ent2}
\end{center}
\end{minipage}
\end{figure}

\subsection{Entropy in quasi-particle approximation}

As an alternative estimate for entropy of the system, we use the
 the quasi-particle (QP) approximation (\ref{eq:entf}) with
the number distribution $n_p(X^0)$  defined in Eq.~(\ref{eq:np}).
The quasi-particle approximation (44) may be reasonable
because the spectral function is nicely peaked \cite{AB} 
near $\omega^2 \sim m^2+{\bf p}^2$
 as seen in Figs.~\ref{fig:sp1} and \ref{fig:sp2}.
Since $n_{\bf p}(X^0)$ is defined locally at time $x^0=y^0$ without the
Fourier transformation, we have no computational difficulty even at
the very early stage, in contrast to the kinetic entropy (\ref{eq:s3}).
We should remark, however, that 
$n_{\bf p}(X^0)$ is here obtained using the solution of
the full KB equation via Eqs.~(\ref{eq:np}) and (\ref{eq:op}).

Although the QP entropy (\ref{eq:entf}) yields somewhat a smaller
value as compared with the kinetic entropy (\ref{eq:s3}),
the evolution profiles of these entropies are
quite similar to each other, except at the early times.
Smaller value for the QP entropy
(\ref{eq:entf}) may be related to the fact that
it neglects the finite width of the spectral distribution over $p^0$.
With the QP approximation we see in Fig.~\ref{fig:ent}
that the entropy production is concentrated at early times 
$mX^0 \lesssim 20$ and slows down at later times $ mX^0 \gtrsim 20$,
approaching an equilibrium value.

In Fig.~\ref{fig:ENTR} we show the ratio between kinetic entropy (40)
and its quasiparticle approximation (44) for 
tsunami and WS initial conditions. 
Until $mX^0\sim 10$
the ratio decreases because of the numerical artifact, explained in the above.
In the range of $ 20 <mX^0 <80$ the ratio becomes constant 
for both initial conditions. 
In the late time region $mX^0>80$ the ratio starts to increase. 
This tendency can be explained from the behavior of time evolution of number density.
In Fig.~\ref{fig:ND} we show the time evolution of number density for tsunami 
and WS initial conditions with $\lambda=4$. At the 
late time $mX^0>80$ with $\lambda=4$ the number density decreases continually. 
In addition the distribution function shown in Figs.~\ref{fig:num} and \ref{fig:num2}
 has the small change for $mX^0>80$. Due to the decrease of number density
and small change for $n_{\bf p}$ the entropy with quasiparticle approximation
(44) increases weakly compared with the kinetic entropy (40).

We compare the evolutions of the QP entropy 
for three values of  the coupling, $\lambda/m^2$=4, 2~and~1 
with the tsunami and WS initial conditions,
respectively, in Figs.~\ref{fig:ent1} and \ref{fig:ent2}. 
In these figures, we see that 
the larger is the coupling the faster is the entropy produced and saturated
to the equilibrium value.
In order to quantify the approach to the equilibrium value,
we fit the entropy evolution 
with a simple functional form
\begin{eqnarray}
s(X^0) = s_{\rm max}- A e^{-\gamma (mX^0)}
\; ,
\label{eq:fit} 
\end{eqnarray}
where $s_{\rm max}$, $A$ and $\gamma$ are parameters.
We chose to fit the evolution in the regions 
$100 \leq mX^0\leq 150$, $300 \leq mX^0 \leq 600$
and $600 \leq mX^0 \leq 900$ for $\lambda$=4, 2 and 1,
because the approach to equilibrium is slower for smaller $\lambda$.
The resultant values for the parameters are summarized in Table.~1.
The parameters $\gamma$  and $s_{\rm max}$ take the same values
independent of the initial conditions
both for the coupling constants $\lambda=$4 and 2.
The $\lambda$ dependence of the parameter $\gamma$ seems
non-trivial, contrary to the
$\lambda^2$ dependence naively inferred from(\ref{htheorem}).  
For $\lambda$=1,
our fit seems still sensitive to the initial conditions.
At the later stage  $\gamma$ might coincide for both initial conditions,
but in this range $mX^0>900$ we have energy errors more than 0.5\%, 
so that we stop our simulation.

\begin{table}
\centerline{
\begin{tabular}{|c|c|ccc|c|ccc|}
\hline
$\lambda$ & $\gamma_0$ & $s_{\rm max}$ & $A$ & $\gamma$ 
          & $\gamma_0$ & $s_{\rm max}$ & $A$ & $\gamma$ \\
\hline
4  &0.24   &2.54  & 0.16& 0.030   &0.14   & 2.54  & 0.14& 0.030    \\ 
2  &0.084   &2.48  & 0.10& 0.0031   &0.036   & 2.44  & 0.13& 0.0031    \\ 
1  &0.027   &2.39  & 0.17& 0.0024   &0.0085   & 2.39  & 0.19& 0.0011    \\ 
\hline
\end{tabular}
}
\caption{Slope parameter $\gamma_0$ near $X^0\sim 0$
and parameters in Eq.~(\ref{eq:fit})
for ``tsunami'' (left) and WS (right) initial conditions.}
\end{table}

\subsection{Particle changing processes}

\begin{figure}[t]
\begin{minipage}{0.45\hsize}
\begin{center}
\includegraphics[height=\textwidth, angle=270  ]{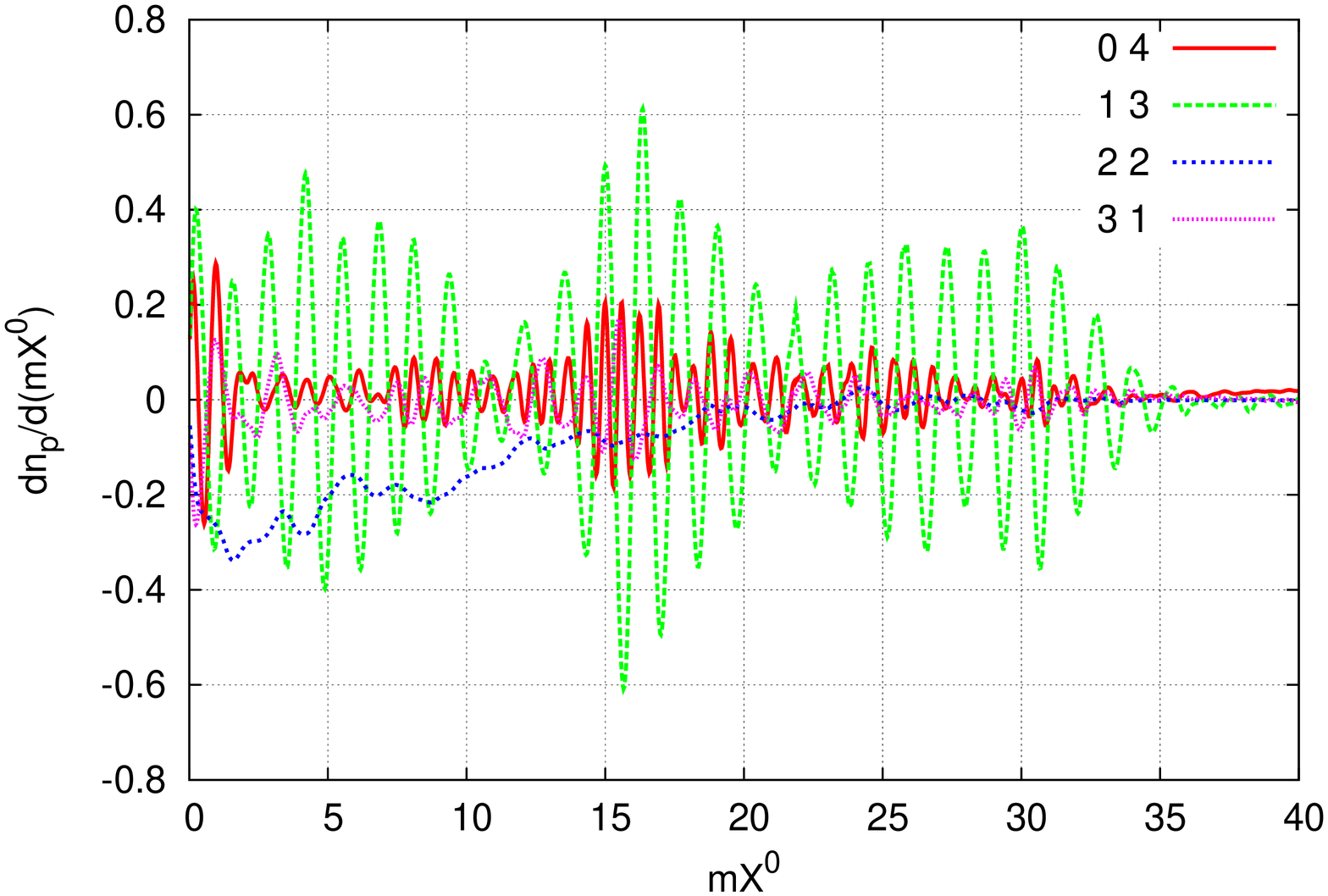}
\caption{Microscopic contributions to 
${dn_{\bf p}}/{dX^0}$ 
at $L \cdot p=7\cdot 2 \pi$ 
as functions of 
$mX^0$ in the case of "tsunami" initial condition with 
$\lambda/m^2=4$. By integrating from $mX^0=0$ to $mX^0=35$ each contribution to
$\delta n_{\bf p}$ is obtained to be 0.51, 0.62, -3.17 and -0.24 for 
0-to-4, 1-to-3, 2-to-2 and 3-to-1 processes.  }

\label{fig:mp}
\end{center}
\end{minipage}
\hfil
\begin{minipage}{0.45\hsize}
\begin{center}
\includegraphics[height=\textwidth, angle=270  ]{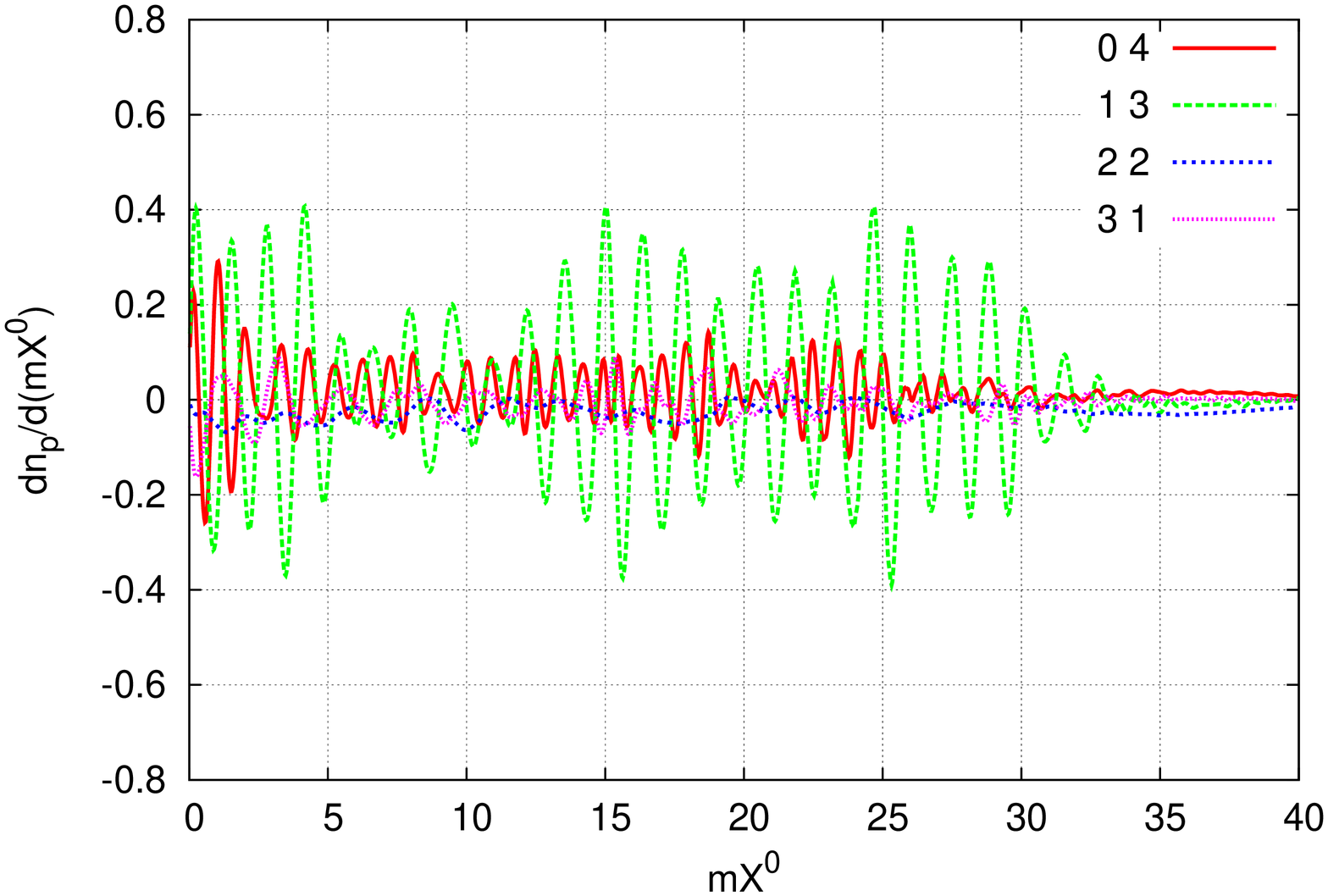}
\caption{
Microscopic contributions to ${dn_{\bf p}  }/{dX^0}$
at $L\cdot p=7\cdot 2 \pi$ as functions of time $mX^0$
in the case of the WS initial condition with $\lambda/m^2=4$. 
By integrating from $mX^0=0$ to $mX^0=35$ each contribution to
$\delta n_{\bf p}$ is obtained to be 0.54, 0.75, -0.86 and -0.26 for 
0-to-4, 1-to-3, 2-to-2 and 3-to-1 processes. }
\label{fig:mp2}
\end{center}
\end{minipage}
\end{figure}

As seen in the behavior of $n(\tilde \omega_{\bf p})$
in Figs.~\ref{fig:num} and \ref{fig:num2}, 
the system evolves toward the equilibrium state. 
The entropy in the QP approximation 
is also written in term of $n(\tilde \omega_{\bf p})$.
Here we study which microscopic process contributes to the change of the
distribution function $n(\tilde \omega_{\bf p})$ in course of the time
evolution. 
Although 
the distribution $n(\tilde \omega_{\bf p})$ in Figs.~\ref{fig:num} and
\ref{fig:num2} is computed using Eqs.~(\ref{eq:np}) and (\ref{eq:op}),
it seems instructive to evaluate the time
derivative $dn_{\bf p}/dX^0$ in the quasi-particle
approximation,
which is given in Eq.~(\ref{eq:mp}) in Appendix \ref{sec:micro}.
Within this approximation we can clearly separate out
the contributions of 0-to-4, 1-to-3, 2-to-2 and 3-to-1 
processes. 
In Figs.~\ref{fig:mp} and \ref{fig:mp2} shown are 
the contributions of each microscopic process 
on the RHS
of $ d n_{\bf p}/d X^0$ Eq.~(\ref{eq:mp}) 
at the momentum $p_n=2 \pi n/L$ with $n=7$
for "tsunami" and WS initial conditions, respectively. 
We set $\lambda/m^2=4$.
The mode with ${\bf p}$ is included as one of the three particles in
the 1-to-3 process while it is formed from three particles in
the 3-to-1 process.

Even in the quasi-particle approximation, 
the number changing processes are possible because of the finite
memory time as was studied in Ref.~\cite{Ikeda2004}.
We see that at early times the number changing processes 
$1\leftrightarrow 3$ contribute as well as $2\leftrightarrow 2$
scattering processes.  
One should recall that $dn_{\bf p}/dX^0=0$ in the Boltzmann
limit in 1+1 dimensions; even the 2-to-2 process is possible only when
it keeps the particle momenta unchanged.

As is seen in Fig.~\ref{fig:num},
the momentum of $n=7$ locates near the peak
for the tsunami initial condition, 
and $n(\tilde \omega_{\bf p})$ decreases
rapidly toward  the equilibrium value. 
In the quasi-particle approximation
shown in Fig.~\ref{fig:mp},
the 2-to-2 process contributes largely to this decrease,
although other number changing processes are also operative at early
times, say $mX^0<35$.
For the WS initial condition, the momentum of $n=7$ 
locates near the shoulder position of the distribution 
and $n(\tilde \omega_{\bf p})$ decreases with time as seen in 
Fig.~\ref{fig:num2}. In this case, the 2-to-2 and other processes
seem equally contributing to $dn(\tilde \omega_{\bf p})/dX^0$.
However, the 2-to-2 contribution is negative and rather non-oscillatory
while others are fast oscillating.

One should bear in mind the limitation of our discussion here.
Namely, these observations are in the quasi-particle approximation,
while we evaluated $n(\tilde \omega_{\bf p})$ using the full solution of the KB
equation which includes both the effects of the finite memory time
and the spectral distribution $\rho(X,p)$. The latter is missed
in the estimate with the quasi-particle approximation.

\section{Discussion}
\label{sec:discussion}

We have studied the time evolution of the $\lambda \phi^4$ theory
in 1+1 dimensions in the framework of the 
KB equations with the statistical and spectral functions,
$F(X,p)$ and $\rho(X,p)$, as basic ingredients.
In this framework
we can take into account two kinds of the ``offshell'' effects;
one is the finite memory time effect and the other is the non-trivial
form of the spectral functions $\rho(X,p)$.

We have seen that the particle number
distribution $n(\tilde \omega_{\bf p})$, which is defined
with the numerical solution of the KB equations, 
converges to the Bose distribution in the time evolution.
We also showed that both the kinetic entropy 
(\ref{eq:s3}) and the QP entropy (\ref{eq:entf}) increase
in course of the evolution. In order to understand 
the mechanism of the entropy production, we analyzed the microscopic
processes within the quasi-particle approximation and found that the
2-to-2 scattering processes as well as the particle number changing
processes are operative in the early stage of the evolution.
It is important to note here that in the Boltzmann description in 1+1
dimensions, the particle distribution function never evolves,
$dn(\tilde \omega_{\bf p})/dX^0=0$, as we mentioned previously.

It is not clear, however, which effect is essential to the system 
thermalization, the finite memory effect or the spectral function
$\rho(X,p)$. It would be interesting to compare the finite memory time
simulations with and without the quasi-particle approximation
in 1+1 dimensions, in order to estimate the importance of the spectral
structure $\rho(X,p)$ in thermalization. We leave this for our future study.

We have introduced formally the kinetic entropy (\ref{eq:s3}) 
associated with the relativistic KB equations,
to the leading order in the gradient expansion.
In general, the gradient expansion applies only when the $X^0$
dependence becomes gentle toward equilibrium, otherwise the higher order 
terms can give substantial effects.
Furthermore, we encountered a difficulty in evaluating the kinetic
entropy (\ref{eq:s3}), the leading term in the gradient expansion, 
because
of the fact that the
spectral function $\rho(X,p)$ shows an oscillation in $p^0$ with the
frequency proportional to $1/X^0$. This is understood as the
uncertainty relation between the energy and the time.
This yields negative values for the occupation number function
$f(X,p)$, and invalidates the expression for the kinetic entropy
(\ref{eq:s3}) at early times. 
This has not been recognized until our attempt for the numerical evaluation.
At sufficiently later times when 
$\rho(X,p)$ and $f(X,p)$ become positive definite,  the resultant
entropy monotonically increases in time.
In contrast,  the QP entropy (\ref{eq:entf}) is expressed in terms of
the number distribution $n_{\bf p}$ obtained from the
two-point function $F$ without the Fourier transformation,
and it shows a monotonic increase in Fig.~\ref{fig:ent}.
Although the QP entropy has no strict foundation in the context of
the KB equation, its time evolution seems physically quite suitable as
an indicator of the system entropy.

We have also analyzed and compared the behavior of kinetic entropy (\ref{eq:s3})
and its quasiparticle approximation (\ref{eq:entf}). 
In the middle range of thermalization the ratio of both entropy becomes constant, so
that entropy (\ref{eq:entf}) is a useful indicator of the thermalization. However 
at the asymptotic stage of thermalization due to the small change of $n_{\bf p}$ 
entropy (\ref{eq:entf}) is affected by the behavior of total number density.
Therefore entropy (\ref{eq:entf}) has difficulty a little  
in estimating asymptotic behavior of the kinetic entropy (\ref{eq:entf}) which
is based on KB equation, but does not take so large difference.

\section{Summary and outlook}
\label{sec:summary}


We have extended the kinetic entropy current to relativistic case from
non-relativistic one on the basis of the KB equation $\lambda \phi^4$ theory. 
The derivation was done in parallel to the non-relativistic case
to the leading order of the gradient approximation
and to the NLO of the skeleton expansion of $\Phi[G]$.
The derived kinetic entropy satisfies the H theorem.

The numerical simulation of the KB equations in 1+1 dimensions has been
performed with the ``tsunami'' and Woods-Saxon initial conditions.
We have seen that the particle number distribution approaches the
Bose-Einstein distribution with time, irrespective of the applied 
initial distributions, which reconfirms the results of  
Ref.~\cite{AB}.
We noted that the Boltzmann equation with 2-to-2 scatterings 
in 1+1 dimensions never thermalizes. Therefore, we see that the
``offshell'' effects included in the KB approach must be very important
for thermalization problem.


In evaluating the kinetic entropy (\ref{eq:s3}) numerically,
we recognized that the Fourier transformation within the limited time
interval for $m X^0 \lesssim 10$,
makes the spectral function $\rho(X,p)$ and
the occupation number function $f(X,p)$ oscillating in $p^0$,
accordingly the expression (\ref{eq:s3}) must become complex-valued.
At later times, on the other hand, the kinetic entropy (\ref{eq:s3})
increases in time, indicating that the system approaches the
equilibrium state. 
Therefore we find that 
the kinetic entropy (\ref{eq:s3}) should be useful when the shape of
the spectral function $\rho(X,p)$ is well resolved, provided that the
gradient expansion is also valid.

As an alternative, we have studied the entropy evolution
in the quasi-particle approximation (\ref{eq:entf}), 
and found that it increases almost monotonically saturating at the
equilibrium value. The late time behavior is similar to that of the
kinetic entropy. We further studied the dependences of the evolution
on the coupling constant and the initial condition.

The KB equations involve the offshell effects of the finite memory time
as well as of the non-trivial spectral functions $\rho(X,p)$ and $F(X,p)$.
We examined effects of the finite memory time 
within the quasi-particle approximation,
and showed that the 2-to-2 process as well as the number changing
processes are operative for producing the entropy at early stage of the
evolution.  
In order to study the importance of the spectral function $\rho(X,p)$,
it will be useful to compare the simulations with and without the
quasi-particle 
approximation in 1+1 dimensions\cite{Ikeda2004}.
We leave this for future study.

The KB equation is one of the most promising approaches to
describe the non-equilibrium processes in the quantum field theories.
Although we limited our simulations in 1+1 dimensions to see the
importance of the ``offshell'' effects, the simulations in high
dimensions are certainly needed, where those effects will be likely
important.
In the field of high-energy heavy-ion collisions, 
early thermalization is one of the important issues under debate.
The KB dynamics may provide a suitable framework to investigate the
early time evolution of such energetic nuclear collisions,
and the entropy introduced in this paper may be one of the useful
quantities to characterize thermalization of the system in course of its
time evolution.  Toward this end, 
extension and practical development to the case of 
gluodynamics are much desired.

\section*{ Acknowledgment}

The author is grateful to Prof.\ H.~Fujii 
for carefully reading the
manuscript and patiently giving much advice for improvements.
He would like to thank also 
Profs.\ T.~Matsui, T.~Biro, K.~Fukushima and T.~Kunihiro for 
fruitful discussions in analytical and numerical calculations 
of non-equilibrium statistical physics.

\appendix

\section{Energy Momentum Tensor}
\label{sec:ene}

We derive the expression for the energy-momentum tensor,
$\Theta^{\mu\nu}$, of the $\lambda \phi^4$ theory. 
The 2PI effective action $\Gamma[G]$ Eq.~(\ref{Action}) 
is invariant
under the translation $x^\mu\rightarrow x^\mu+\epsilon^\mu$.
Following Noether's procedure, we apply the
position dependent translation
$x^\mu\rightarrow x^\mu+\epsilon^\mu(x)$ to compute
the change of the action
$\delta\Gamma =
\int_x \partial_\nu \left [
\epsilon_\mu(x) \Theta^{\mu \nu}(x)
\right]$.
For $\epsilon_\nu$ independent of $x$, 
we can prove the current conservation 
$\partial_\nu \Theta^{\mu\nu}=0$
as a result of the invariance of the action $\delta \Gamma=0$. 
The energy-momentum tensor $\Theta^{\mu\nu}$ reads from $\delta \Gamma$
as the coefficient factor of 
$\partial_\nu \epsilon_\mu(x)$\cite{{Baym},{IKV},{AST}}.

Under the position dependent translation,
Green's function changes as
\begin{eqnarray}
G(x,y) \rightarrow G'(x,y) 
\equiv
 G(x+\epsilon(x), y+\epsilon(y))=
G(x,y)+ \epsilon^\lambda (x) \partial^x_\lambda G(x,y)
+ \epsilon^\lambda(y) \partial^y_\lambda G(x,y).
\end{eqnarray} 
Then the change of each term in the action (\ref{Action}) is
calculated as follows:
the first term in Eq.~(\ref{Action})
leaves no $\epsilon$ term 
\begin{eqnarray}
\delta \left[\frac{i}{2} {\rm Tr}\ln G^{-1} \right]
&=&
-\frac{i}{2} {\rm Tr} \frac{1}{G} \delta G
=
-\frac{i}{2}\int _{x,y} G^{-1}(x,y) 
\left[\epsilon^\mu (y) \partial_\mu ^y G(y,x)
+\epsilon^\mu (x)\partial_\mu ^x G(y,x)
\right] 
\nonumber \\
&=& -i \int _x \epsilon^\mu(x) \partial_\mu^x \delta(x-x) =0
\; . 
\end{eqnarray}
The second term gives rise to
\begin{eqnarray}
\delta \left[\frac{i}{2} G_0^{-1} G \right] 
&=&
 -\frac{i}{2} \int_{x,y} \left[ \left(\partial_x^2+m^2 \right)
\delta(x-y) \right]
\delta G(x,y)
\nonumber \\
&=&-\frac{1}{2} \int _x \epsilon^\mu (x) \partial_x^\nu \Big [ \delta(x-y)
 \partial_\mu^x\partial_\nu^y \left( G(x,y)+G(y,x)
\right) 
\nonumber \\
&&
-\delta (x-y)g_{\mu \nu} \partial_x^\lambda \partial_\lambda^y G(x,y) 
+m^2g_{\mu \nu} \delta(x-y) G(x,y) \Big ]
\; ,
\end{eqnarray}
where we have used
\begin{eqnarray}
\int _{x,y} \partial_x^\mu 
\left[ \delta(x-y)G(x,y)\right] 
= \int_{x,y} \delta(x-y) 
\left[ \partial_x^\mu G(y,x)+ \partial_x^\mu G(x,y) \right] \; .
\end{eqnarray}
The third term  yields
\begin{eqnarray}
\delta \left[ \frac{1}{2} \Phi[G] \right] 
= \frac{1}{2} \int _{x,y} \frac{\delta \Phi }{\delta G(x,y)} 
\left(\epsilon^\mu (x)\partial_\mu^x G(x,y)+
 \partial^\mu (y)\partial_\mu^y G(x,y) \right).
\end{eqnarray}
The third term 
$\delta \left[\frac{1}{2}\Phi[G] \right]$
can be rewritten more conveniently by use of the Jacobian.
For example the tadpole part in $\Phi[G]$ changes
under the translation as
\begin{eqnarray}
\delta \left[ \lambda\int _x G(x,x)^2 \right]
 = 2\lambda \int_x  G(x,x) 
\left (
\epsilon^\mu (x) \partial_\mu^x G(x,y) +
\epsilon(y)^\mu\partial_\mu^y G(x,y) 
\right).
\end{eqnarray} 
However we observe that the translation can be dealt with the change of
variables in the integral as
\begin{eqnarray}
\lambda \int_x G(x+\epsilon(x),\ x+\epsilon(x))^2&=& 
\int_{x'} {\rm det} \left(1+ \frac{\partial x^\mu}{\partial x'^\nu} \right) 
G(x',\ x')^2   
\end{eqnarray}
and therefore the change can be recast to
\begin{eqnarray}
\delta \left[ \lambda\int _x G(x,\ x)^2 \right]&=&
 -\lambda \int_{x'} \frac{\partial^\mu \epsilon}{\partial x^\mu}\
G(x',x')^2
= -\lambda \int_{x} \frac{\partial^\mu \epsilon}{\partial x^\mu}\
G(x,x)^2
\; .
\end{eqnarray}
In this way, the total change of $\Phi[G]$ can be re-expressed by
using the Jacobian.
Furthermore, the number of integration coincides with the
power of the coupling $\lambda$ in general. This means that
the Jacobians under the translation can be absorbed in
the change of the coupling $\lambda \rightarrow \lambda \zeta (x)$
and that $\delta \Phi$ can be rewritten as \cite{{Baym},{IKV},{AST}}
\begin{eqnarray}
\delta \left[\frac{1}{2} \Phi \right] 
= \int_{x}
\partial _\mu  \epsilon(x) ^\mu 
\frac{\delta \Phi}{\delta \zeta(x)}\Big | _{\zeta=1}.
\end{eqnarray}

Since 
$\delta \Gamma= \int_x \partial_\nu \epsilon_\mu (x)
\Theta^{\mu \nu}(x)$, 
the energy-momentum tensor is found as
\begin{eqnarray} 
\Theta^{\mu \nu}(x) &=& \frac{1}{2} \int_y 
\Big[
 \delta (x-y)\partial^\mu_x \partial^\nu _y 
            \left( G(x,y)+G(y,x) \right) 
-\delta (x-y) g^{\mu \nu} 
            \partial_x^\lambda \partial^y_\lambda G(x,y) 
\nonumber \\
&&+g^{\mu \nu} \delta(x-y)m^2 G(x,y) 
\Big]
-\frac{1}{2} g^{\mu \nu} 
 \frac{\delta \Phi}{\delta \zeta(x)} \Big |_{\zeta=1}
\end{eqnarray}

In the case of uniform space,
by taking the Fourier transform with respect to the
spatial relative coordinate,
we obtain the explicit expressions for
the energy $\Theta^{00}$ and the pressure $\Theta^{11}$ 
as%
\footnote{The counter term in energy and pressure is the same.}:
\begin{eqnarray}
\Theta^{00}(X^0) \!\!\!\!&=&\!\!\!\!
 \frac{1}{2} \int \!\! \frac{d^d p}{(2\pi)^d} 
({\bf p}^2+m^2 + \partial_{x^0} \partial_{y^0}   )
 F(x^0,y^0; {\bf p}) \Big | _{x^0=y^0=X^0} 
-\frac{1}{2} \frac{\delta  \Phi }{\delta \zeta} -
\text{counter term}  \; ,
\\
\Theta^{11}(X^0) \!\!\!\!&=&\!\!\!\!
 \frac{1}{2} \int \!\! \frac{d^d p}{(2\pi)^d} \left(
{\bf p}^2-m^2 + \partial_{x^0} \partial_{y^0}   
\right) 
F(x^0,y^0; {\bf p}) \Big | _{x^0=y^0=X^0} 
+\frac{1}{2} \frac{\delta  \Phi }{\delta \zeta} 
+\text{counter term} \; .
\end{eqnarray}
Here the counter term cancels out the divergence in
$\frac{\delta  \Phi}{\delta \zeta}$. In $1+1$ dimensions
the divergence is only in the tadpole part which can be in the similar
manner to that of \cite{AB}.  

In the NLO in $\lambda$  of the skeleton expansion,
the tadpole and sunset diagrams contribute to
$\delta  \Phi /\delta \zeta$. 
Then the total energy and pressure are decomposed as
\begin{subequations}
\begin{align}
E_{\rm tot}(X^0)&= E_{\rm kin}(X^0)+ E_{\rm tad}(X^0)+E_{\rm sun}(X^0) 
\; ,
\label{eq:ecomp} \\
P_{\rm tot}(X^0)&= P_{\rm kin}(X^0)+ P_{\rm tad}(X^0)+P_{\rm sun}(X^0)
\; ,
\label{eq:Pcomp}
\end{align}
\end{subequations}
where
\begin{subequations}
\begin{align}
E_{\rm kin}(X^0)&=
 \frac{1}{2} \int \!\! \frac{d^d p}{(2\pi)^d} 
({\bf p}^2+m^2 + \partial_{x^0} \partial_{y^0})
 F(x^0,y^0; {\bf p}) | _{x^0=y^0=X^0}  
\; ,
\label{eq:ekin}
\\
P_{\rm kin}(X^0)&=
\frac{1}{2} \int \!\! \frac{d^d p}{(2\pi)^d} 
({\bf p}^2-m^2 + \partial_{x^0} \partial_{y^0}   
) 
F(x^0,y^0; {\bf p}) | _{x^0=y^0=X^0}
\; ,
\label{eq:pkin}
\\
E_{\rm tad}(X^0)&=
-P_{\rm tad}(X^0)= 
\frac{1}{4} \int \!\!\frac{d^dp}{(2\pi)^d} \Sigma_{\rm tad}^{}(X^0)
F(X^0,X^0;{\mathbf p}) 
+\frac{1}{2} \int \!\! \frac{d^dp}{(2\pi)^d} 
\delta m_{\rm tad}^2\;  F(X^0,X^0; {\bf p}) 
\; ,
\label{eq:etad}
\\
E_{\rm sun}(X^0)&
=-P_{\rm sun}(X^0)=
\frac{1}{4} \int \!\! \frac{d^dp}{(2\pi)^d} I(X^0, {\bf p})
\; .
\label{eq:esun}
\end{align}
\label{energycontent}
\end{subequations}
Here $\delta m_{\rm tad}^2$ denotes the mass counter term:
\begin{equation}
\delta m_{\rm tad}^2 = -\frac{\lambda}{2} \int \!\! \frac{d^dp}{(2\pi)^d}
 \frac{1}{2\omega_{\bf p}}
\; ,
\end{equation}
and 
\begin{eqnarray}
I(X^0,{\bf p}) &= \int _0^{X^0} \!\! dt' \left[ \Sigma_\rho(X^0,t';{\bf p}) F(t',X^0;{\bf p}) -
 \Sigma_F(X^0,t';{\bf p}) \rho (t',X^0;{\bf p}) \right] 
\end{eqnarray}
is finite in 1+1 dimensions
\footnote{In the case of $3+1$ dimensions, the renormalization
of self-consistent theories is required and developed in \cite{HK}.}.

\section{Microscopic processes in the Kadanoff-Baym equation}
\label{sec:micro}

The evolution of the particle number distribution can be understood
in terms of the microscopic processes in the quasi-particle
approximation.
See \cite{{BergesReview},{Ikeda2004}} for a similar discussion.

First, we differentiate the distribution $n_{\bf p}$ 
defined in Eq.~(\ref{eq:np})
with respect to the time $t$ to find
\begin{eqnarray}
\left(\frac{1}{2}+n_{\bf p} \right)\partial_t n_{\bf p} &=&
 \int_{t_0}^{t}\!\! dz^0
\Big \{ \left [ \Sigma_\rho(t,z^0;{\bf p}) F(z^0,t;{\bf p}) 
-\Sigma_F(t,z^0; {\bf p} )\rho(z^0, t; {\bf p}) \right]
\partial_t F(t,t';{\bf p})|_{t=t'} 
\nonumber \\
&&-\left[ \Sigma_\rho(t,z^0;{\bf p})\partial_t F(z^0,t;{\bf p}) 
-\Sigma_F(t,z^0; {\bf p} ) \partial_t\rho(z^0, t; {\bf p})
 \right] F(t,t;{\bf p}) \Big \}.
\label{eq:dnpdt}
\end{eqnarray}
Meaning of the memory integrals on the RHS
of Eq.~(\ref{eq:dnpdt})
becomes clear if we take the quasi-particle limit
for $\rho$ and $F$ as given in Eqs.~(\ref{eq:rfree}) and (\ref{eq:ffree}).
We find  
$\rho(t,t';{\bf p}) =\tilde \omega({\bf p})^{-1}\sin[( \tilde \omega({\bf p})(t-t')]$
and
$F(t,t';{\bf p})=
\tilde \omega({\bf p})^{-1} \cos [\tilde \omega({\bf p}) (t-t')]
(n_{\bf p}+\tfrac{1}{2} )$.
We substitute these to the self-energy
$\Sigma_\rho$ and $\Sigma_F$ at the NLO 
in Eq.~(\ref{eq:dnpdt}) 
and 
arrive at the following expression:
\begin{eqnarray}
\partial_t n_{\bf p}(t) &=&
\frac{\lambda^2}{3} \int \! \frac{d^d {\bf q}}{(2\pi)^d} 
\frac{d^d {\bf k}}{(2\pi)^d} \int_{t_0}^t \!\!dt'
\frac{1}{2 \tilde \omega({\bf p}) 2 \tilde \omega({\bf q}) 
         2 \tilde \omega({\bf k}) 2 \tilde \omega({\bf p-k-q}) }
\nonumber \\
&&
 \Big \{  \left[ ( 1+ n_{\bf p} ) ( 1+ n_{\bf q})
                 ( 1+n_{\bf k}) (1+ n_{\bf p-k-q})
            -n_{\bf p} n_{\bf q} n_{\bf k} n_{\bf p-k-q}(t')
         \right]
\nonumber \\
&\times & \cos \left[\left(  \tilde \omega({\bf p})+
\tilde \omega({\bf q})+\tilde  \omega({\bf k})+\tilde  \omega({\bf p-k-q }) \right)(t-t') \right]
\nonumber \\
&&+
3  \left[ ( 1+ n_{\bf p} ) ( 1+ n_{\bf q})
         ( 1+n_{\bf k}) n_{\bf p-k-q}-n_{\bf p} n_{\bf q} n_{\bf k}
         ( 1+n_{\bf p-k-q}) (t') \right]
\nonumber \\
& \times & \cos \left[\left(\tilde \omega({\bf p})+
\tilde \omega({\bf q})+\tilde  \omega({\bf k})-\tilde  \omega({\bf p-k-q }) \right)(t-t') \right]
\nonumber \\
&&+
3  \left[ ( 1+ n_{\bf p} ) ( 1+ n_{\bf q})n_{\bf k} n_{\bf p-k-q}
        -n_{\bf p} n_{\bf q} (1+n_{\bf k})(1+ n_{\bf p-k-q})(t') \right]
\nonumber \\
&\times & \cos \left[\left(
\tilde \omega({\bf p})+\tilde \omega({\bf q})-\tilde  \omega({\bf k})-\tilde  \omega({\bf p-k-q }) \right)(t-t') \right]
\nonumber \\
&&+
 \left[ ( 1+ n_{\bf p} )  n_{\bf q}n_{\bf k} n_{\bf p-k-q}
       -n_{\bf p} (1+n_{\bf q}) (1+n_{\bf k})(1+ 
n_{\bf p-k-q})(t') \right]
\nonumber \\
&\times  & \cos \left[\left(
\tilde \omega({\bf p})-\tilde \omega({\bf q})-\tilde  \omega({\bf k})-\tilde  \omega({\bf p-k-q }) \right)(t-t') \right]
\Big \}.
\label{eq:mp}
\end{eqnarray}  
We can interpret physically each microscopic process in the bracket
$\{ \cdots \}$ of Eq.~(\ref{eq:mp}).
The first term represents creation and annihilation
processes of four particles $0 \leftrightarrow  4$. 
The second term describes the process $ 1 \rightarrow  3$ where
$n_{\bf p}$ is involved as one of 
the three particles and its reverse.
The third term corresponds to $2\leftrightarrow 2$ scattering process.
The last term describes the
decay of $n_{\bf p}$ to 3 particles and its reverse.

The number changing processes are allowed because of 
the finite memory time $t-t'$;
the energy conservation in each microscopic process can be violated
\cite{Ikeda2004}. 
We evaluated these contributions and showed then in Figs.~\ref{fig:mp}
and \ref{fig:mp2}. Oscillatory behaviors seen in these figures
obviously come from $\cos$ functions. 
Removing the memory time effect by taking the infinite time 
limit $t-t_0 \rightarrow \infty$, we recover the strict energy
conservation in the microscopic process, 
$\lim_{t-t_0 \rightarrow \infty}
\int_{t_0}^t dt' \cos(\omega(t-t')) =
\pi \delta(\omega)$. In this limit only the $2\leftrightarrow 2$
scatterings are allowed
and then we obtain the Boltzmann equation (\ref{Bol}).


\begin{thebibliography}{99}
\bibitem{BergesReview}
J. Berges, AIP Conf. Proc. {\bf 739}, 3 (2005) [hep-ph/0409233].

\bibitem{Noneq}
For example, see 
`{\it `Progress in Nonequilibrium Green's Functions III},'' 
M. Bonitz and A. Filinov (eds.), J. Phys. Conf. {\bf 35}, 1 (2006).


\bibitem{HeinzK}
U. W. Heinz, AIP Conf. Proc. {\bf 739}, 163 (2005). 



\bibitem{BMSS}
R. Baier, A. H. Mueller, D. Schiff and D. T. Son, Phys. Lett. 
{\bf B502}, 51 (2001).

\bibitem{Weibel}
For review, S.\ Mrowczynski, PoS CPOD2006: 042 (2006)
[hep-ph/0611067].




\bibitem{Mrow1993}
S. Mrowczynski,  Phys.\ Lett.\ {\bf B314}, 118 (1993);
  Phys.\ Rev.\  C {\bf 49}, 2191 (1994);
  Phys.\ Lett.\  B {\bf 393}, 26 (1997).


\bibitem{Arnold2003}
  P.~Arnold, J.~Lenaghan and G.~D.~Moore,
  JHEP {\bf 0308}, 002 (2003).

\bibitem{Rebhan2005}
  A.~Rebhan, P.~Romatschke and M.~Strickland,
  Phys.\ Rev.\ Lett.\  {\bf 94}, 102303 (2005).
\bibitem{Dumitru2007}
  A.~Dumitru, Y.~Nara and M.~Strickland,
  Phys.\ Rev.\  D {\bf 75}, 025016 (2007).

\bibitem{Romat2006}
  P.~Romatschke and R.~Venugopalan,
  Phys.\ Rev.\  D {\bf 74}, 045011 (2006).




\bibitem{NielsenO}
A.\ Iwazaki, Phys.\ Rev.\ C {\bf 77}, 034907 (2008).\\
H.\ Fujii and K.\ Itakura, Nucl.\ Phys.\ A {\bf 809}, 88 (2008).

\bibitem{XuG2007}
Z. Xu and C. Greiner, Phys.\ Rev.\  C {\bf 71}, 064901 (2005);
 Phys. Rev. C {\bf 76}, 024911 (2007).



\bibitem{LW}
J. Luttinger and J. Ward, Phys. Rev. {\bf 118}, 1417 (1960).

\bibitem{BK}
G. Baym and L. Kadanoff, Phys. Rev. {\bf 124}, 287 (1961).

\bibitem{Baym}
G. Baym, Phys. Rev. {\bf 127}, 1391 (1962).
\bibitem{KB62}
L.P. Kadanoff, G. Baym, {\it Quantum Statistical Mechanics}
   (Benjamin, New York, 1962).
\bibitem{CJT}
J. M. Cornwall, R. Jackiw and E. Tomboulis, Phys. Rev. D {\bf 10}, 2428 (1974)
\bibitem{Schwinger}
J. Schwinger, J. Math. Phys. {\bf 2} (1961) 407.



 \bibitem{Keldysh}
L.V. Keldysh, ZHETF {\bf 47} 1515 (1964) [Sov. Phys. JETP {\bf 20}, 235 (1965)].


\bibitem{Danielewicz}
P. Danielewicz, Ann. Phys. (N.Y.) {\bf 152}, 305 (1984)

\bibitem{AB}
G. Aarts and J. Berges, Phys. Rev. D {\bf 64}, 105010 (2001)

\bibitem{BC}
J. Berges and J. Cox, Phys .Lett. B {\bf 517}, 369 (2001)


\bibitem{JCG}
S. Juchem, W. Cassing and C. Greiner, Phys. Rev. D {\bf 69}, 025006 (2004) 
\bibitem{AST}
A. Arrizabalaga, J. Smit and A. Tranberg, Phys. Rev. D {\bf 72}, 025014 (2005)
\bibitem{LM}
M. Lindner and M. M. M\"uller; Phys. Rev. D 73 125002 (2006)
\bibitem{Berges}
J. Berges, Nucl. Phys. {\bf A699}, 847 (2002) 

\bibitem{IKV4}
Y.B. Ivanov, J. Knoll, and D.N. Voskresensky, Nucl. Phys. {\bf A672}, 313 (2000)
\bibitem{Kita}
T. Kita, J. Phys. Soc. Jpn. {\bf 75}, 114005 (2006)
\bibitem{IKV}
Y.B. Ivanov, J. Knoll, and D.N. Voskresensky, Nucl. Phys. {\bf A657}, 413 (1999)
\bibitem{IKV2}
Y.B. Ivanov, J. Knoll, and D.N. Voskresensky, Ann. Phys. {\bf 293}, 126 (2001)

\bibitem{CH}
E. Calzetta, B.L. Hu, Phys. Rev. D {\bf 37} (1988) 2878.
\bibitem{Cassing} 
W. Cassing, nucl-th 08080715
\bibitem{BM}
W. Botermans and R. Malfliet, Phys. Rep. 198 (1990) 115 
\bibitem{IMGM}
I. Montvay and G. M\"unster; Quantum Fields on a Lattice, Cambridge University Press (1994)
\bibitem{RM}
J. Rau and B. M\"uller, Phys. Rep. 272 (1996) 1

\bibitem{HK}
H. van Hees and J. Knoll, Phys. Rev. D {\bf 65}, 025010 (2002);Phys. Rev. D {\bf 65}, 105005 (2002);
Phys. Rev. D {\bf 66}, 025028 (2002)
\bibitem{Ikeda2004}
T. Ikeda, Phys.\ Rev.\ D {\bf 69}, 105018 (2004).





\end{thebibliography}
\end{document}